\documentclass[11pt]{article}
\usepackage{graphics,epsfig} 
\begin{document}

\title{Timing and Counting Precision in the Blowfly
Visual System}
\date{}
\author{Rob de Ruyter van Steveninck and 
William Bialek \\NEC Research Institute,\\ 4 Independence Way,\\ 
Princeton NJ 08540,\\USA}

\maketitle

\begin{abstract}

We measure the reliability of signals at three levels
within the blowfly visual system, and present a
theoretical framework for analyzing the experimental
results, starting from the Poisson process. We find that
blowfly photoreceptors, up to frequencies of 50-100~Hz
and photon capture rates of up to about
$3\cdot10^5$/s, operate well within an order of 
magnitude from ideal photon counters. Photoreceptors 
signals are transmitted to LMCs 
through an array of chemical synapses. We quantify a
lower bound on LMC reliability, which in turn provides a
lower bound on synaptic vesicle release rate, assuming
Poisson statistics. This bound is much higher than what
is found in published direct measurements of vesicle
release rates in goldfish bipolar cells, suggesting that
release statistics may be significantly
sub-Poisson. Finally we study H1, a motion sensitive
tangential cell in the fly's lobula plate, which
transmits information about a continuous signal by
sequences of action potentials. In an experiment with
naturalistic motion stimuli performed on a
sunny day outside in the field, H1 transmits information
at about 50\% coding efficiency down to millisecond 
spike timing precision. Comparing the measured 
reliability of H1's response to motion steps with the 
bounds on the accuracy of motion computation set by 
photoreceptor noise, we find that the fly's brain makes 
efficient use of the information available in the 
photoreceptor array.

\end{abstract}  

\section{Introduction}

\noindent Sensory information processing plays a crucial
role in the life of animals, including man, and perhaps
because it is so important it seems to happen without
much  effort.
In contrast to this, our subjective experience suggests that
activities of much lower urgency, such as proving mathematical theorems
or
playing chess, require substantial conscious mental 
energy, and this seems to make them inherently 
difficult. This may deceive us into thinking that 
processing sensory information must be trivially easy. 
However, abstract tasks such as those mentioned are now
routinely performed by computers, whereas the problems
involved in making real-life perceptual judgments are
still far from understood. Playing chess may
seem much more difficult than discerning a tree in a
landscape, but that may just mean that we are very efficient at
identifying trees. It does not tell us anything
about the ``intrinsic'' difficulty of either of the two
tasks.

Thus, to find interesting
examples of information processing by the brain we do
not need to study animals capable of abstract thinking.
It is sufficient that they are just good at processing
sensory data. Partly for this reason, sensory
information processing by insects has been an active
field of study for many years. Undeniably, insects in
general have simpler brains than vertebrates, but
equally undeniably, they do a very good job with what
they do have. Noting
that insect brains are very small, Roeder (1998)
remarks:
\begin{quote}
Yet insects must compete diversely for their
survival against larger animals more copiously equipped.
They must see, smell, taste, hear, and feel. They must
fly, jump, swim, crawl, run, and walk. They must sense
as acutely and act as rapidly as their predators and
competitors, yet this must be done with only a fraction
of their nerve cells.
\end{quote}
Spectacular examples of
insect behavior can be found in Tinbergen (1984) and
Berenbaum (1995). Brodsky (1994) describes 
the acrobatics of fly flight: ``The house fly can stop
instantly in mid flight, hover, turn itself around its
longitudinal body axis, fly with its legs up, loop the
loop, turn a somersault, and sit down on the ceiling,
all in a fraction of a second.'' An example of some 
of this performance
is shown in Fig.~1, and clearly the acrobatic behavior 
displayed there must be mediated by impressive feats of 
sensory information processing.

\begin{figure}[tbp]
\begin{center}
\includegraphics[keepaspectratio,width=1.0\textwidth]
{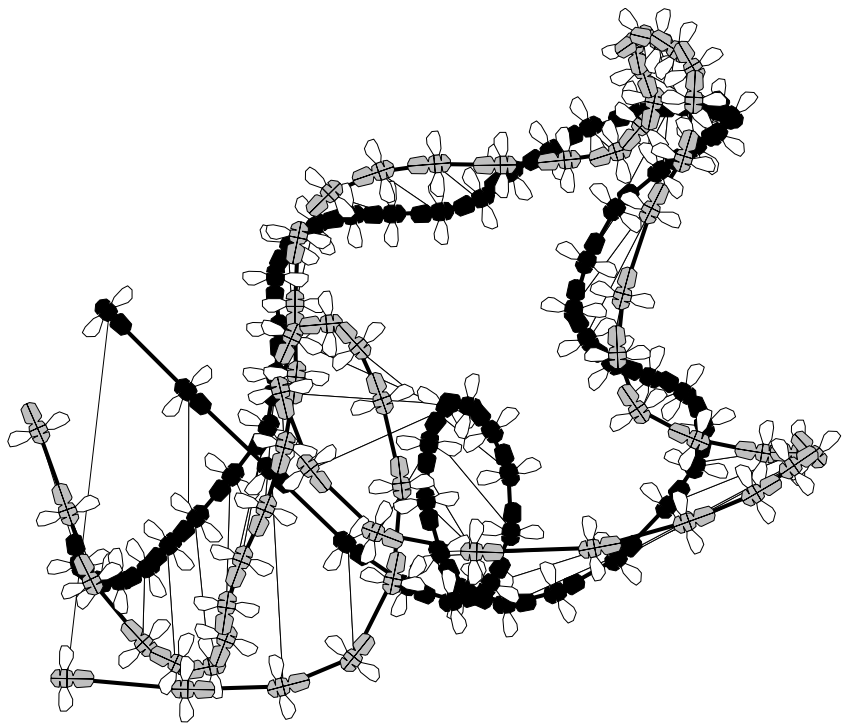}
\caption[ ]{\footnotesize
  Two flies in a chase. The black fly is chased
  by the gray one for a duration of about a second. Flies
  (\emph{Fannia canicularis}) were filmed from below, and
  their positions estimated every 20 ms. The chasing
  fly estimates the position of the leading one purely
  by means of visual input. Modified from Fig.~4 in
  Land and Collett (1974).}
\end{center}
\end{figure}

In this chapter we will look at the tip of this iceberg,
and study some aspects of visual information processing
in the blowfly. Insects lend themselves well for
electrophysiological and behavioral studies, especially
for quantitative analysis, and the emphasis will be on
quantifying the \emph{accuracy} of neural signals and
neural computations, as advocated explicitly by
Bullock (1970). This is a fundamentally
probabilistic outlook, requiring a suitable statistical
description of signal and noise. It also is the
principal description from the point of view of
representation and processing of information. For
example, if you (or the brain) are asked to estimate the
light intensity in the outside world based on the
reading of a photoreceptor voltage, then of course you
need to know the gain, i.e., the conversion factor from
light intensity to photoreceptor voltage. To find the
light intensity you just divide the voltage by this
gain. The accuracy of your estimate depends on the
magnitude of the cell's voltage noise, and this
translates into an uncertainty in light intensity
through the same gain factor. But then the uncertainty
in the estimate depends on the ratio of the signal
voltage and the noise voltage, and not on the specific
value of the gain.  Note that the gain is a ``biological''
parameter---why does the cell produce millivolts
and not hundreds of millivolts?---while the accuracy of 
estimates is measured in the same units as the physical stimulus 
in the outside world.  Further, we will see that there
are often absolute limits to accuracy that are set by physical
principles.  Thus by studying the accuracy of estimates we
move from a characterization of neurons that is specific
to the their biological context toward a characterization
that is more absolute and universal.

The example of estimating light intensity by looking at the voltage
in a photoreceptor admittedly is very simple.
Nonetheless, it illustrates an important point, namely
that we can quantify on an absolute scale the performance of neurons as
encoders and processors of sensory information
even without a complete understanding of the
underlying physiological mechanisms.  The challenge
that we address in later sections of this chapter is
to give a similarly quantitative description for neurons
at higher stages of processing, where the interpretation
of the neural signals is much more difficult.

We begin at the beginning of vision, namely light.
Because light is absorbed as an irregular stream of
photons, it is an inherently noisy physical signal. This
is modeled mathematically by the Poisson process, of
which we give some mathematical details which will be
used later on. Then we will focus on the performance of
the photoreceptor cells of the fly's retina which
capture light and convert it into an electrical signal.
These cells encode signals as graded, or analog,
variations of their membrane potential. This
representation is noisy in part due to the fluctuations
in the photon flux, and in part due to limitations
imposed by the cell itself, and we can tease  apart these
contributions. Then we will look at the accuracy
of chemical synaptic transmission between the
photoreceptor cell and the LMC (Large Monopolar
Cell) \index{Large Monopolar Cell} by comparing signal
and noise in the presynaptic and postsynaptic cells. A
chemical synapse releases discrete vesicles, and
therefore the Poisson process is a natural starting
point for a mathematical description of signal transfer
across the synapse.

Of course, having a representation of light intensities
in the outside world does not in itself constitute
vision. It is the task of the brain to \emph{make  sense}
of the ever fluctuating signals in the photoreceptor
array. As an example we consider the estimation of
wide-field motion from signals in the photoreceptor
array, which is a relatively simple neural computation.
We will analyze the limits to the reliability of this
computation, given the reliability of the photoreceptor
signals, and we will compare this to the reliability of
performance of H1, a spiking, wide field
motion sensitive cell in the fly's brain.

By comparing the measured reliability of cells with the
physical limits to reliability set by the noise in the
input signal we get an idea of the statistical
efficiency of nerve cells and of neural computation. This
also makes it possible to quantify rates of information
transfer, which, although by no means the whole story,
nevertheless captures the performance of nerve cells in
a useful single number.

\section{Signal, Noise and Information Transmission in a
Modulated Poisson Process}

A Poisson process \index{Poisson process} is a sequence
of point events with the property that each event occurs
independently of all the others. We will treat some
of the mathematics associated with Poisson
processes and shot noise, but our emphasis is on an
intuitive, rather than a rigorous presentation.
Many of the issues discussed here can be
found in a treatment by Rice (1944, 1945), which also
appears in an excellent collection of classic
papers on noise and stochastic processes by Wax (1954).
A more comprehensive mathematical treatment is given by
Snyder and Miller (1991).

The Poisson process is a useful model first of all
because it describes the statistics of photon capture
very well. There are exotic light sources which deviate
from this (Saleh and Teich 1991), but
biological organisms deal with normal thermal light
sources. The photoreceptor response can then be modeled
to first approximation as shot noise, which is a
linearly filtered Poisson process. The filtering process
itself may be noisy, as is the case
in phototransduction. Consequently, we
also treat the more general case in which the filter
itself is a random waveform. The Poisson process
is often used to model other point processes
occurring in the nervous system, such as vesicle release
at a chemical synapse, or trains of action potentials.
We will present examples of both, and show that
for vesicle release the Poisson process
is probably not a good approximation, and for
describing spikes in H1 it is certainly
not good. But even then it is useful to understand the
Poisson process, because it provides a simple example
with which to ground our intuition.

\subsection{Description of the Poisson process}
\label{POISSDESC}
\begin{figure}[tbp]
\begin{center}
\includegraphics[keepaspectratio, 
 width=1.0\textwidth]{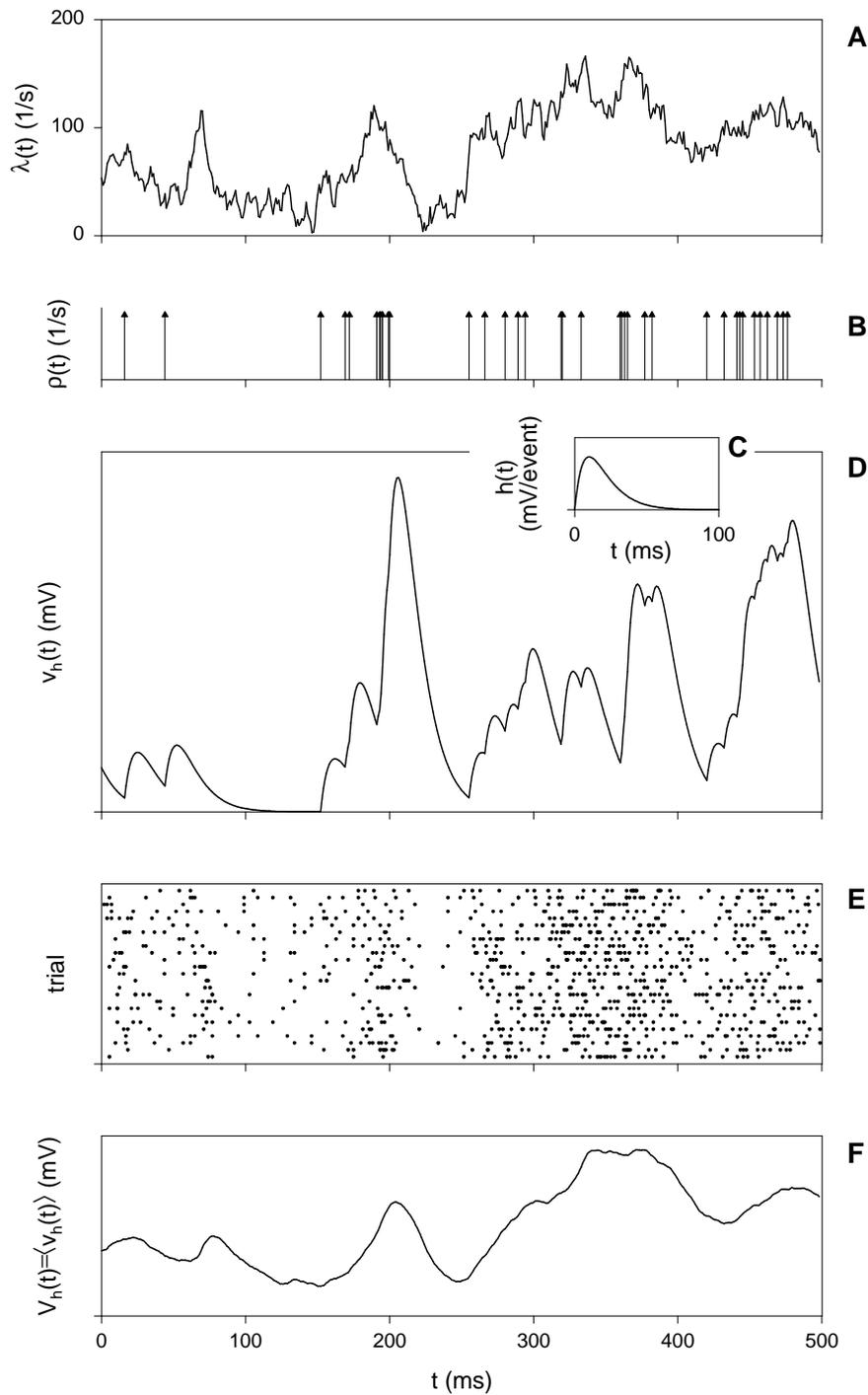}
\caption[ ]{\footnotesize Illustration of some of the
  basic concepts used in the analysis. A: Time
  dependent rate, $\lambda(t)$, of a Poisson process. B: 
  A single realization of statistically independent
  events on the time axis, generated from the rate
  function shown in A. The arrows represent delta
  functions. C: A temporal filter, $h(t)$, used to
  smooth the train of events in B to produce the
  example of a single shot noise trace, $v_h(t)$, shown
  in panel D. E: A raster representation of
  the outcome of 25 trials in which events are generated
  by the rate function $\lambda(t)$ shown in A. F: 
  The average trace, $V_h(t)=\langle v_h(t)\rangle$,
  of a large number of trials filtered by $h(t)$.}
\end{center}
\end{figure}

\noindent Many of the notions treated in this section
are illustrated in Fig. 2. Fig. 2B for example, shows
a single realization of a point process on the time
axis. This can be written as a series of events
occurring at times $t_1,t_2,\cdots$. A useful
mathematical abstraction is to represent each event as a
delta function, $\delta(t-t_k)$, so that the full series
is a sum of these functions:

\begin{equation}
\label{eq:TRAIN}
\rho(t)=\sum_k \delta(t-t_k).
\end{equation}

\noindent Here the delta function has the dimension of
inverse time, and an area equal to 1, so that the
integral of $\rho(t)$ over a time window will be equal
to the number of events in that window. What
distinguishes a Poisson process from other sequences of
events is that the occurrence of each event is random
and completely independent of the occurrence of all the
others. Eqn. \ref{eq:TRAIN} describes one single
outcome, which is  analogous to observing the number on
a die in one single throw. Although the particular
result may be important in a game of
dice, it is not the particular
realization that we are interested in here. Instead we
wish to derive what we can expect under
``typical'' conditions. A construction that helps in such
derivations is to think not of one single outcome of the
process, but of a great number of independent
realizations. This is similar to the concept in
statistical mechanics of an ensemble \index{ensemble} of
independent systems, each in its own state, and all
obeying the same macroscopic conditions (see for example
Schr\"{o}dinger 1952). The advantage of this mental
picture is that it provides a convenient starting point
for reasoning about typical or average behavior.

For the
die we usually take it for granted that the chance of
getting any particular number from 1 to 6 is just 1/6.
But we could also imagine an immense number of dice
thrown at random, and if we were to count the ones that
landed on 6, we would find their proportion to the total
to be very close to 1/6. Obviously, using the ensemble
concept for reasoning about dice is a bit overdone.
Later on we will look at experiments where the same time
dependent stimulus sequence is repeated a large number
of times, while neural responses are recorded. In that
case it will be useful to think about a large set of
outcomes---our independent trials---and to distinguish
ensemble averages \index{average!ensemble} over that
set of trials from time averages \index{average!time}.
Ensemble averages will generally be indicated by
$\langle \cdots \rangle$, and time averages by
$\overline {\cdots}$. Here we use the term ``ensemble'' in
a loose sense, and primarily to make the distinction
with time averages. A synthesized example 
is shown in Fig. 2E,
while Fig. 17B shows a set of spike timing data from a 
real experiment.

Suppose that Eqn. \ref{eq:TRAIN} represents a
realization of a Poisson process that has a constant,
time independent rate of $\lambda$\index{Poisson
process!rate}, that is on average we expect to see
$\lambda$ pulses in a one second time window.
How does this relate to the sequence described by
Eqn.~\ref{eq:TRAIN}? In the ensemble picture we construct
a large number, $M$, of independent realizations of
$\rho(t)$, denoted
$\rho_1(t),\rho_2(t),\cdots,\rho_M(t)$, with total
counts $N_1,N_2,\cdots,N_M$. The ensemble average,
$r(t)$ of $\rho(t)$ is then:

\begin{equation}
\label{eq:TRAIN_AVG}
r(t)=\langle\rho(t)\rangle=\lim_{M\to\infty}
\frac{1}{M} \sum_{m=1}^M \sum_{k_m=1}^{N_m}
\delta(t-t_{k_m})=\lambda.
\end{equation}

\noindent
Obviously $\lambda$ has the same dimension, inverse
time, as the delta function.
But how do we derive this result in the first place? Let
us introduce a rectangular pulse, $[\Pi(t/\Delta t)] /
\Delta t$. Here $\Pi(t/\Delta t)$ is by definition equal
to 1 for $|t|<\Delta t /2$ and zero otherwise
(see Bracewell, 1978, and Rieke et al. 1997, A.1), so
that $[\Pi(t/\Delta t)] /\Delta t$ is $\Delta
t$ wide, and $1/\Delta t$ high. This obviously has unit
area, just like the ``real'' delta pulse, and it becomes
more like a delta pulse if we let $\Delta t$ shrink to
zero. If $\Delta t \ll 1/\lambda$, then such pulses will
(almost) never overlap. If we now also imagine time to
be discretized in bins $\Delta t$, then it becomes a
matter of bookkeeping to count what proportion of the
realizations in the ensemble has a pulse in a given time
bin. The answer is $P_1=\Delta t \cdot \lambda$. To
normalize to units of time we must divide this by the
binwidth, $\Delta t$, so the rate will be $\lambda$.

Thus we expect to
see about $\mu=\lambda \cdot T$ events occurring in a
window of $T$ seconds in a single realization,
independent of where we put the window. Of course, $\mu$
is an average, and we do not expect to observe exactly
$\mu$ events each time, if only because $\mu$ can be
any nonnegative number, while each event count value
must be an integer. It is interesting to know how the
values in each trial are distributed. This is described
by the Poisson distribution\index{Poisson distribution}:

\begin{equation}
\label{eq:POISSONDIST2}
P(N)=\frac{(\lambda T)^N}{N!} \cdot
e^{-\lambda T}.
\end{equation}

\noindent It is easy to show that this distribution has
mean value $\mu = \langle N \rangle=\lambda \cdot T$,
and variance $\sigma^2=\langle N^2-\langle N \rangle^2
\rangle$. Further,
$\langle N^2 - \langle N \rangle ^2
\rangle =\langle N \rangle$, and this means that
$\sigma^2 = \mu$, which is an important property of the
Poisson distribution. A distribution which is more narrow
than a Poisson distribution has $\sigma^2 < \mu$, and is
often called sub-Poisson. The broader distribution,
with $\sigma^2 > \mu$, is
called super-Poisson.

\subsection{The modulated Poisson process}
\label{sect:MPP}

\index{Poisson process!modulated} If we want to study
visual information transfer we should not confine
ourselves to looking at steady light levels, as constant
visual stimuli have a zero rate of information
transmission. One reading of a signal value can of
course carry information, but if it does not change over
time the information rate, in bits/s, goes to zero.
Perhaps for that reason the visual system does not seem
to care very much for constant input signal levels.
Later we will see the example of Large Monopolar Cells
which filter out practically all of the standing signal
in the fly's photoreceptors. In real life, signals are
changing all the time, and in real experiments we
usually modulate signals and probe the response of nerve
cells to these changes. To study the response of
photoreceptors one may deliver flashes of light, sine
waves of various frequencies, pseudorandom signals, and
so on. All of these are really modulations of an
underlying Poisson process, namely the rate at which
photons are captured. We therefore must extend our
description of the homogeneous Poisson process to an
inhomogeneous, or modulated, version in which the rate
is a function of time. So suppose that the rate is time
dependent $\lambda=\lambda(t)$, as depicted in Fig. 2A.
If we have a large number of outcomes of a Poisson
process modulated with $\lambda(t)$, then in analogy to
Eqn. \ref{eq:TRAIN_AVG} the ensemble average, $r(t)$,
must equal $\lambda(t)$:

\begin{equation}
\label{eq:MODTRAIN_AVG}
r(t)=\langle\rho(t)\rangle=\lim_{M\to\infty}
\frac{1}{M} \sum_{m=1}^M \sum_{k_m=1}^{N_m}
\delta(t-t_{k_m})=\lambda(t).
\end{equation}

\subsection{Correlation functions and spectra}
\label{POISS_FREQ}

\index{correlation function} \index{spectrum}
Because the pulses in a Poisson process occur
independently of one another it is easy to compute
correlation functions and power spectral
densities.   These correlation functions are
natural objects from a physical point of view,
and are used widely in the analysis of experimental
data.  There is a difficulty, however, in that
correlation functions are averages, and we have two
different notions of averaging:  averaging over
time, and averaging over
an ensemble of responses to identical inputs.
In many ways the situation is similar to that in
the statistical mechanics of disordered systems,
where we can average over thermal fluctuations,
over different realizations of the quenched
disorder, or over both.  Our discussion here and in the next 
section is brief, summarizing results that are largely well 
known. What we aim at ultimately is the derivation of 
Eqn.~\ref{eq:RATE_EST}, connecting 
observable quantities to the underlying rates
of Poisson events.

We recall that if we average the signal $\rho(t)$ over
many trials in which the inputs are the same, then we
obtain the Poisson rate $\lambda(t)$, $\langle \rho(t)
\rangle = \lambda(t)$. Since different pulses occur
independently of one another, once we know the rate
$\lambda(t)$ there is no additional mechanism to carry
correlations across time. On the other hand, since
$\rho(t)$ is built out of delta functions at the moments
of the point events, there will be singularities if we
compute correlations at short times. It is
straightforward to show that 

\begin{equation}
\langle \rho(t) \rho(t')\rangle
= \lambda(t) \lambda (t') + \lambda(t) \delta (t-t' ).
\end{equation}

\noindent Thus if we compute the ``connected'' correlation
function---subtracting off the means of the terms taken 
separately---all that remains is the delta function term:

\begin{eqnarray}
\langle \rho(t) \rho(t')\rangle_c
&\equiv&
\langle \rho(t) \rho(t')\rangle
-
\langle \rho(t) \rangle\langle 
\rho(t')\rangle\nonumber\\ &=&
\lambda(t) \delta (t-t' ) .
\end{eqnarray}

\noindent It is interesting that if we now average over
time, the answer is insensitive to the time dependence
of the rate. More precisely, the time average of the
connected correlation is 

\begin{equation}
\overline{\langle \rho(t) \rho(t+\tau)\rangle}_c \equiv 
\lim_{T\to\infty} \frac{1}{T}\int_0^{\infty} dt~ 
\langle \rho(t) \rho(t+\tau)\rangle = 
\overline{\lambda} \delta(\tau) ,
\end{equation}

\noindent and this is true whether or not the Poisson
process is modulated. In this sense the fluctuations in
the Poisson stream are independent of the modulations,
and this is an important way of testing for the validity
of a Possion description. 

Instead of averaging over an ensemble of trials we can
also average directly over time. Now we have no way of
distinguishing between true correlations among
successive events (which would indicate a departure from
Poisson statistics) and correlations that are ``carried''
by the time dependence of the rate itself. The result is
that

\begin{eqnarray}
\overline{[\rho(t)\rho(t+\tau)]} &=&
\overline{[\lambda(t)\lambda(t+\tau)]} + {\bar \lambda}
\delta (\tau),
\\
\overline{[\rho(t)\rho(t+\tau)]}_c
&=&
\overline{[\Delta \lambda (t) \Delta \lambda(t+\tau)]}+
{\bar \lambda} \delta (\tau),
\label{timeavgrhocorr}
\end{eqnarray}

\noindent where the fluctuations in the rate are defined
as $\Delta \lambda (t) = \lambda (t) - 
\overline{\lambda}$. Thus the (connected) 
autocorrelation of the pulse train, computed by 
averaging over time, has a contribution from the 
(connected) autocorrelation function of the rate and a 
delta function contribution.

We recall that the integral of $\rho(t)$ over a time
window counts the number of events in that window. The
variance of the count found in a window thus is equal to
a double integral of the $\rho$---$\rho$ correlation
function (see, for example, Appendix 2 in Rieke et al.
1997). The delta function in the autocorrelation leads
to a term in the count variance that grows linearly with
the size of the window over which we count, much as a
delta function in the velocity autocorrelation for a
Brownian particle leads to diffusion; here we have a
diffusion of the count. Since the mean count grows
linearly with the size of the counting window, the
variance and mean are growing in proportion to one
another, and in fact the proportionality constant is
one. This equality of variance and mean is one of the
defining features of the Poisson process. 

If we try to analyze the signal $\rho(t)$ in the
frequency domain, then the first step is to subtract the
time average and define $\delta\rho(t) = \rho(t) -
\overline{\rho(t)}$. Note that $\overline{\rho(t)} =
\overline{\lambda}$ if we average for a sufficiently 
long time. Then if we Fourier transform and compute the 
obvious correlation functions, we obtain: 

\begin{eqnarray}
\delta{\hat\rho}(f) &=& \int dt \exp(+2 \pi i f t) \delta\rho(t),
\label{spec1}\\
\langle \delta{\hat\rho}(f) \delta{\hat\rho}(f ')\rangle
&=& \delta (f + f' ) S_{\delta\rho}(f),
\label{spec2}\\
S_{\delta\rho}(f) &=& S_{\Delta \lambda} (f) + 
\overline{\lambda}, \label{spec3}
\end{eqnarray}

\noindent where we use $\hat{\cdot}$ to denote Fourier
transforms. Thus the spectral density of fluctuations in
$\rho$, $S_{\delta \rho}$, is related to the spectral 
density of rate fluctuations, $S_{\Delta \lambda}$, 
plus a ``white noise'' term with an amplitude equal to 
the mean rate. This structure parallels that for the 
correlation functions in Eqn. (\ref{timeavgrhocorr}) 
because the spectra and correlation functions form a 
Fourier transform pair,

\begin{equation}
\overline{[\rho(t)\rho(t')]}_c = \int df
\exp[-2\pi i f (t-t')]S_{\delta\rho}(\omega );
\end{equation}

\noindent this is the Wiener--Khinchine theorem.

Notice that if we have an ensemble of many trials with
the same input, then we can define a different
``fluctuation'' in the signal $\rho$ by subtracting off
the time dependent mean $\langle \rho(t) \rangle =
\lambda(t)$. Thus if we write $\Delta \rho(t) = \rho(t)
- \lambda(t)$, then the equations analogous to 
(\ref{spec2}) and (\ref{spec3}) above are as follows:

\begin{eqnarray}
\langle \Delta{\hat\rho}(f) \Delta{\hat\rho}(f')\rangle
&=&  \delta (f + f' ) S_{\Delta\rho}(f )\\
S_{\Delta\rho}(f ) &=& {\bar \lambda}.
\end{eqnarray}

\noindent Again we see that the connected fluctuations
are independent of the modulation. 

\subsection{Shot noise}
\label{SHOTNOISE}

\index{shot noise}
Idealized Poisson events have zero
duration, or equivalently, infinite bandwidth. This is
clearly unrealistic; real signals have finite rise
and decay times. We can give the Poisson process some
more meat by replacing every zero-duration event with a
fixed waveform and having all the waveforms in the
train be additive. Another way of saying this is that
the train of events in Eqn.~\ref{eq:TRAIN} is filtered by
a linear filter $h(t)$ \index{linear filter} (see Fig.
2C), and this is often a fair first order model for the
physics underlying shot noise, such as pulses of
electrons from a photomultiplier tube. The output is
found by convolving \index{convolution} this filter with
the sequence of delta functions:

\begin{equation}
\label{eq:SHOTNOISE}
v_{h}(t)=\sum_k \delta(t-t_k) \otimes h(t)
= \sum_k h(t-t_k),
\end{equation}

\noindent
where we use $v_{h}$ to denote the result,
because in our applications the filtered
process will always be a voltage. An example of a
filtered process is given in Fig. 2D. If the width of
$h(t)$ is very small compared to $1 / \lambda$, then the
shot noise still looks a lot like the original Poisson
process in the sense that there are clearly separated
events. If, however, $h(t)$ is much larger than the
average separation between events then $v_h(t)$ will be
a rather smooth signal, and its value will be
distributed according to a Gaussian. This is easy to
see: When at any one point in time, $v_h$ is the sum of
a large number of filtered events, spread randomly in
time, the central limit theorem (Feller 1966) tells us
that the probability distribution of that sum approaches
a Gaussian, at least for reasonable forms of $h(t)$.

\index{power spectral density!shot noise}
\index{shot noise!power spectral density}
Convolution in the time domain is equivalent to
multiplication of the Fourier transforms in the
frequency domain, and so, using Eqn.~\ref{spec3}, we can
write down the power spectral density of the shot noise
process as:

\begin{equation}
\label{eq:PSPDENS_S}
S_v(f)
=\left[S_{\Delta \lambda}(f)
+ \overline{\lambda}\right]\cdot |\hat{h}(f)|^2,
\end{equation}

\noindent
where we use the subscript $v$ in $S_v$ to
remind us that the noise power density will
represent the power density spectrum of voltage fluctuations
measured from a cell.  Notice that the spectral density 
includes contributions from the ``signal''---the time 
variation of $\lambda$---as well as from noise due to 
randomness in the event times. Below we will isolate the
``true'' noise component of the spectrum, which we will 
call $N_v(f)$. 

Again these results are what we obtain by making a
Fourier analysis, which is related to the correlation
functions that are defined by averaging over time. On
the other hand, if we can observe many responses to the
same inputs, then we can compute an average over this
ensemble of trials, to define the \index{Poisson
process!modulated and filtered} ensemble averaged
output, $V_h(t)$: 

\begin{eqnarray}
\label{eq:FILTMODTRAIN}
V_h(t) \equiv \langle v_h(t)\rangle  & = &
\lim_{M\to\infty} \frac{1}{M}
\sum_{m=1}^M \sum_{k_m=1}^{N_m}
\left[\delta(t-t_{k_m})
 \otimes h(t)\right] \nonumber \\
 & = & \left[ \lim_{M\to\infty}\frac{1}{M}\sum_{m=1}^M
\sum_{k_m=1}^{N_m} \delta(t-t_{k_m}) \right] \otimes
h(t) \nonumber \\  & = & \lambda(t) \otimes h(t),
\end{eqnarray}

\noindent
where the first step follows because
convolution is a linear operation, so that the order of
summation and convolution can be changed, and the second
step is from Eqn. \ref{eq:MODTRAIN_AVG}. The end
result is illustrated in
Fig. 2F. Equation \ref{eq:FILTMODTRAIN} thus simply
states that the ensemble averaged output of a filtered
modulated Poisson process is equal to the filtered rate.
This will be used in analyzing the transduction of
contrast by photoreceptors and second order cells, which
for moderate contrast fluctuations are reasonably close
to linear. Thus from an experiment where we generate
$\lambda(t)$ and measure $V_h(t)$ we
can in principle derive $h(t)$.

It will be convenient to
write $\lambda(t)$ as the product of a constant,
$\lambda_0$, and a contrast modulation function $c(t)$:
$\lambda(t)=\lambda_0 \cdot [1+c(t)]$, where $c(t)$
represents contrast as a function of time. This also
conforms closely to the experiment, where we use $c(t)$
as a signal to modulate a light source with an average
photon flux equal to $\lambda_0$. In the experiment we
will present the same stimulus waveform a large number
of times to generate an ensemble of responses. Now,
using the frequency representation of
Eqn.~\ref{eq:FILTMODTRAIN} we get:

\begin{eqnarray}
\label{eq:TRF}
\hat{V}_h(f) &=&
\hat{\lambda}(f)\cdot \hat{h}(f) \nonumber \\
&=&
\lambda_0 \cdot \hat{c}(f)\cdot \hat{h}(f)  \ \ \ \ \ \
\ (f \ne 0). \end{eqnarray}

\noindent
Thus the Fourier transform
of the ensemble averaged response equals the product of
the Fourier transforms of the rate and the
filter. In an experiment we set the
stimulus, $\lambda(t)$ in the above equation, and we
measure $V_h(t)$. Using these data and
Eqn.~\ref{eq:TRF} we can directly compute the transfer
function $\hat{h}(f)$ that translates the stimulus into
the average response. But we can also look at the fluctuations around
this average,

\begin{equation}
\Delta v_h(t)=v_{h}(t)-V_h(t),
\end{equation}

\index{fluctuations!power
spectral density}\index{power spectral density!of
fluctuations}
\noindent
and now we want to compute the power spectral density of these
fluctuations.   The key is to realize that

\begin{equation}
\Delta v_h(t) = h(t) \otimes \Delta\rho(t) ,
\end{equation}

\noindent which means that

\begin{eqnarray}
\label{eq:SHOTNOISEPD} 
N_{v} (f ) &=& | \hat{h}(f) |^2 S_{\Delta r} (f) 
\nonumber
\\ &=& \lambda_0 | \hat{h}(f) |^2 ,
\end{eqnarray}

\noindent where we use the notation $N_v$ because this 
truly is a spectral density of noise in the response 
$v_h$. Again the crucial result is that the noise  
spectral density is determined by the mean counting 
rate, independent of the modulation.  There is even a
generalization to the case where the filter
$h(t)$ is itself random, as discussed in the next section.

These results
give us a way of calculating the underlying average
Poisson rate from the observed quantities in the
experiment:

\begin{equation}
\label{eq:RATE_EST}
\lambda_0=\frac{|\hat{V}_h(f)|^2}
{N_v(f) \cdot |\hat{c}(f)|^2}.
\end{equation}

\noindent This of course is valid only for an ideal,
noiseless, system, the only noise being that introduced
by Poisson fluctuations. We could choose to
apply Eqn.~\ref{eq:RATE_EST} to the quantities we
measure, and then consider the result as a measurement
of an \emph{effective} Poisson rate \index{Poisson
rate!effective} (see also Dodge et al. 1968). This can
then be compared to the rate we expect from an
independent measurement of photon flux (which we can
make---see the experimental sections) to assess how close
the real system comes to being ideal. But we can make
the decription a bit more realistic by introducing a
random filter in the chain of events leading to the
cell's response.
 
\subsection{A Poisson process filtered by a
random filter}
\label{RANDFILT}

\index{Poisson process!and random filter}\index{shot
noise!and random filter}
The analysis presented above
relies on the filter $h(t)$ having a prescribed, fixed
shape. That assumption is not entirely realistic. More
to the point, we are interested in characterizing the
limitations of the system as a transmitter of
information, and so it is precisely this deviation from
strict determinacy that we wish to analyze.
Phototransduction is a biochemical process, rooted in
random collisions of molecules. Not surprisingly
therefore, quantum bumps in photoreceptors are known to
fluctuate (Fuortes and Yeandle 1964, Baylor et al. 1979,
Wong et al. 1980, Laughlin and Lillywhite 1982), varying
both in shape and in latency time. We would like to
incorporate the effect of such fluctuations in our
formulation, and we do that here for the simplest case.
Suppose a Poisson event at time $t_k$ is filtered
by a filter $h_k(t)$, that the shape of this filter is
drawn from a probability distribution of filter shapes,
$\mathcal{P}[h(t)]$, and that these draws are
independent for different times $t_k$. As before, the
contributions of the filtered events are assumed to add
linearly. We can picture this distribution of filter
shapes again as an ensemble, and of course this ensemble
has some average filter shape, $\langle h(t) \rangle$.
Because everything is still linear, we can exchange
orders of averaging, and so it is easy to see that we
can replace the fixed shape $h(t)$ in Eqn.~\ref{eq:TRF}
by its ensemble average. In other words in the case of
independently fluctuating filters we obtain:

\begin{equation} \label{eq:TRF_FLUCT}
\hat{V}_h(f)=\langle
\hat{v}_h(f)\rangle = \hat{\lambda}(f) \cdot \langle
\hat{h}(f)\rangle,
\end{equation}

\noindent and instead of being able to measure the fixed
shape of a filter we have to settle for characterizing
its average.

Finally we should compute the effect of variable filter
shapes on the power density spectrum, that is we want
the analogue of Eqn.~\ref{eq:SHOTNOISEPD}. Here it 
is useful to remember that the power density spectrum is
the Fourier transform of the autocorrelation function.
If bumps of variable shapes are generated at
random times, then each bump correlates with itself.
The correlations with the others, because
their shapes and arrival times are assumed independent,
will lead to a constant. The autocorrelation of
the bump train is then the ensemble average of the
autocorrelation of all individual bumps. Likewise the
power density spectrum is the ensemble average of the
power spectra of the individual bumps. The end result
is then that:

\begin{eqnarray}
N_v(f) &=& \lambda_0 \cdot \langle |\hat{h}(f)|^2
\rangle \label{eq:COMBIN_FLUCT_1} \\
|\hat{V}_h(f)|^2 &=& \lambda_0^2 \cdot
 |\langle \hat{h}(f)\rangle|^2 \cdot |\hat{c}(f)|^2.
\label{eq:COMBIN_FLUCT_2}
\end{eqnarray}

\noindent
We can now define  an effective Poisson rate
\index{Poisson rate!effective} that we compute from the
experimental data:

\begin{eqnarray}
\label{eq:RATE_FLUCT_EST}
\hat{\lambda}_{\mathrm{eff}}(f) &=&
\frac{|\hat{V}_h(f)|^2} {N_v(f) \cdot  |\hat{c}(f)|^2}
\nonumber \\  &=& \lambda_0 \cdot \frac{|\langle
\hat{h}(f)\rangle |^2}  {\langle |\hat{h}(f)|^2
\rangle}, \end{eqnarray}

\noindent
and this is in general a function of frequency,
because when $h(t)$ fluctuates in shape, the behavior of
$\langle |\hat{h}(f)|^2 \rangle$ is different from that
of  $|\langle\hat{h}(f) \rangle|^2$.

It is worthwhile fleshing out what effect fluctuations
in different bump characteristics have on
$\hat{\lambda}_{\mathrm{eff}}(f)$. Here we consider
variations in amplitude and in latency. There is good
evidence that such variations in bump parameters
occur almost
independent of one another (Howard 1983, Keiper and
Schnakenberg 1984). Variations in the amplitude of the
bump \index{bump!amplitude variations} can be modeled
by assuming that individual bump shapes are described by
$\beta \cdot h_0(t)$, where $\beta$ is a random variable
with mean $\langle\beta\rangle=1$, and variance
$\sigma^2_{\beta}$, and where $h_0(t)$ has a fixed
shape. It is easy to derive that in that case we have
$\hat{\lambda}_{\mathrm{eff}}(f)=\lambda_0 /
(1+\sigma^2_{\beta})$. In other words, random variations
in bump amplitude lead to a frequency independent
decrease in the effective Poisson rate. Any form of
noise that leads to a spectrally flat effective decrease
in photon flux is therefore sometimes referred to as
multiplicative (Lillywhite and Laughlin 1979). An
important special case is that of a decrease in the
photon flux. This can be described by taking $\beta$ to
be a random variable with value either 0 or $1/p_1$,
with $p_1$ the probability of a photon being transduced
(so that $\langle\beta\rangle=1$, as required, and
$\sigma^2_{\beta}=1/p_1 -1$). This leads to
$\hat{\lambda}_{\mathrm{eff}}(f)=p_1\cdot\lambda_0$,
as expected.

Another important source of
randomness is a fluctuating latency time from photon
absorption to bump production.
\index{bump!latency fluctuations} Suppose
that we have a fixed shape, $h_0(t)$ as before, but that
the time delay is distributed, independent for different
bumps, according to $p(t_{\mathrm{lat}})$. Displacing
random events in a random way preserves the independence
and the mean rate, so that there is no effect on the
noise power density. The timing with respect to external
modulations is compromised, however, and this mostly
affects the reliability at high frequencies. Again,
starting from Eqn.~\ref{eq:RATE_FLUCT_EST} it is easy to
derive: $\hat{\lambda}_{\mathrm{eff}}(f)=\lambda_0 \cdot
|\hat{p}(f)|^2$---that is, the frequency dependence of
the effective Poisson rate is given by the Fourier power
transform of the latency distribution (de Ruyter van
Steveninck and Laughlin 1996b). Because
$p(t_{\mathrm{lat}})$ is a probability distribution, its
Fourier transform must go to 1 for $f \to 0$, and will
go to 0 for $f \to \infty$ if the distribution is
smooth. Thus, if bumps
have random latencies the effective Poisson rate will be
frequency dependent. Low frequencies are not affected
whereas the effective rate goes to zero in the limit of
very high frequencies.

Fluctuations in the duration of bumps as well as
external additive noise in general have frequency
dependent effects on $\hat{\lambda}_{\mathrm{eff}}(f)$.
If the aim is to describe the phototransduction
cascade and the other processes occurring in the
cell, then it is interesting to try and tease all the
contributions apart, and this may not be easy. Here,
however, our goal is more modest, in that we want to
quantify the reliability of photoreceptors and the LMCs onto 
which they project; compare the results to the limits imposed by 
the stochastic nature of light; and explore some of the 
consequences of photoreceptor signal quality for visual 
information processing.

\subsection{Contrast transfer function and equivalent
contrast noise}
\label{CTRF_ECN}

It is useful to define two other quantities. The
first is the contrast transfer function,
\index{contrast!transfer function} defined by:

\begin{equation}
\label{eq:CTRF}
\hat{H}(f)=\frac{\hat{V}_h(f)}{\hat{c}(f)} =
\lambda_0 \cdot \langle \hat{h}(f) \rangle,
\end{equation}

\noindent which expresses the cell's gain not as the
translation from single photon to voltage, but from
contrast to voltage. This is practical because
photoreceptors and LMCs work mostly  in a regime of light
intensities where bumps are fused, and then it is
natural to think of these cells as converting contrast
into voltage. In addition to this transduced contrast
there is voltage noise. It is conceptually useful to
express these noise fluctuations as an effective noise $\eta_c$
added to the contrast itself,

\begin{equation}
{\hat v}(f) = \hat{H}(f)[{\hat c}(f) + {\hat\eta}_c(f)] .
\end{equation}

\noindent The spectral density of this effective noise source is then
the
equivalent contrast noise power density\index{equivalent contrast noise
power} $N_c(f)$, and has units of (contrast)$^2$/Hz.  Since contrast
itself is dimensionless this means that $N_c$ has units of 1/Hz, and
hence $1/N_c$ has the units of a rate; we will see below that for an
ideal photon counter this rate is exactly the mean rate at which photons
are being counted. To find $N_c(f)$ we inverse filter the measured
voltage noise power density by the contrast power transfer function:

\begin{equation}
\label{eq:CNPD}
N_c(f)=\frac{N_v(f)}{|\hat{H}(f)|^2},
\end{equation}

\noindent
and we can easily derive that
$N_c(f)=1/\lambda_{\mathrm{eff}}(f)$. If we now have a
cascade of elements, such as the photoreceptor and the
LMC, and we measure the equivalent contrast noise power
density at each stage, we would like to define the
accuracy of the interposing element, in this case the
array of synapses between photoreceptors and LMC. Using
the equivalent contrast noise power density it is easy
to do this: If we measure $N_{c1}(f)$ and $N_{c2}(f)$
for two elements in a cascade, then for all $f$ we must
have $N_{c2}(f) \ge N_{c1}(f)$, and the difference is
the contribution of the element in between. In the
particular case of photoreceptors and LMCs we have to be
careful to include the effect of having six
photoreceptors in parallel. This, assuming we may treat
them as statistically independent but otherwise equal,
is easy to do: When elements are combined in parallel we
divide $N_c(f)$ for the individual one by the number of
elements to get the equivalent contrast noise power
density of the combination. The charm of working with
the equivalent contrast noise power density is that it
allows us to compute the result of
combining elements in series and in parallel, in the
same way that we calculate the net resistance of series
and parallel combinations of resistors.

\section{ The Early Stages of Fly Vision}

\subsection{Anatomy of the blowfly visual system}

A good proportion of the surface of a fly's head is
taken up by its compound
eyes\index{fly!anatomy}\index{compound eye}, which is a
direct indication that the eyes are very important to
the fly. This is also clear from other considerations:
The energy consumed by all the photoreceptors in the
blowfly's retina in bright light is about 8$\%$ of
the total energy consumption of a fly at rest (Laughlin
et al. 1998). In the blowfly, each eye has about 5000
lenses, corresponding to 5000 pixels of visual input.
These pixels are distributed over almost a hemisphere,
so the two eyes provide the fly with almost complete
surround vision. The male's eyes are somewhat larger
than the female's, because the visual fields of the male's
two eyes overlap in the dorsofrontal region of visual
space. This region is used in detecting and chasing
females.

\subsection{Optics}
\label{Optics}

\begin{figure}[tbp]
\begin{center}
\includegraphics[keepaspectratio,width=0.8\textwidth]
{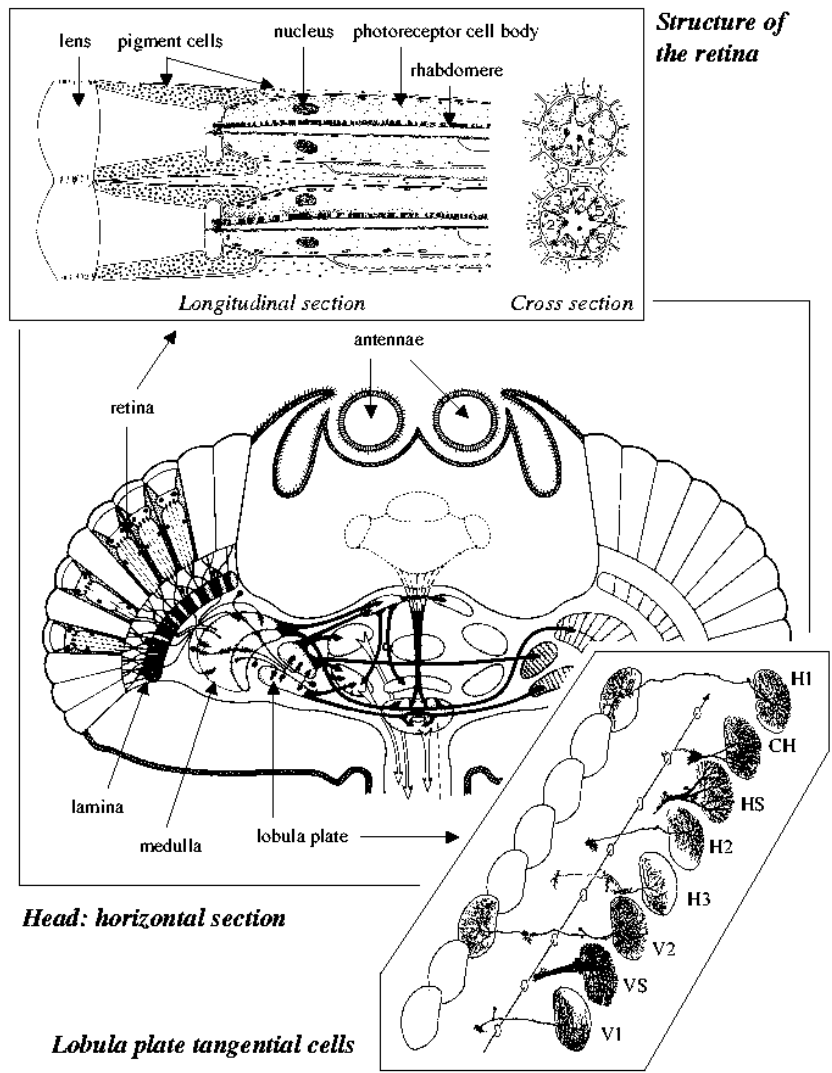}
\newpage
\caption[ ]{\footnotesize Top: Part of the retina, modified
  from Stavenga (1974), showing two ommatidia, each
  containing one lens, several pigment cells, and 8
  photoreceptor cells. Light enters the lens and is
  focused on the tips of the rhabdomeres. It then travels
  in a bound mode along the rhabdomere, in which it can be
  absorbed by a membrane bound rhodopsin molecule (Stavenga 
  1995). Through a series of biochemical steps this then 
  leads to a measureable electrical response across the cell
  membrane. Center: Schematic horizontal cross section
  through the head of a typical fly, modified from
  Kirschfeld (1979). The areas mainly relevant for visual
  information processing are the retina with its
  photoreceptors, the lamina, where photoreceptor signals
  are combined and filtered, and the medulla and lobula
  complex, where more complex information processing takes
  place. Bottom: Exploded view of giant tangential cells
  of the lobula plate, modified from Hausen (1981). The
  outlines of the two lobula plates and the esophagus
  are drawn in thin lines. The lobula plate is an output
  station of the visual system, where tangential cells
  collect motion information over large areas of the
  visual field. From here signals are sent to the thoracic
  ganglion where they are used in the control of flight.
  Cells drawn in this figure collect their information in
  the right lobula plate; H1 and V2 project to the
  contralateral plate. The tangential cells of the lobula
  plate encode information about wide field motion. All
  are direction selective, that is their firing rate
  increases for motion in their preferred direction, while
  spike activity is suppressed for motion in the opposite
  direction. The H cells code horizontal, and the V cells
  vertical motion. CH has a more complicated directional
  selectivity. The labels HS and VS refer to groups of
  cells. The tangential cells are unique and identifiable,
  so that they can be numbered. H1 in particular is a good
  cell to record from both because it is very easy to find
  on the contralateral side, and because it is inward
  sensitive, responding preferentially to motion from the
  side toward the midline of the animal. The combination
  of this directional selectivity with the contralateral
  projection is unique, so that H1 can be identified 
  unambiguously from the electrophysiological recording.}
\end{center}
\end{figure}

\noindent The compound eye of insects forms an image
onto an array of photoreceptors through a large number
of tiny lenses (with diameters ranging typically from 10
to 30~$\mu\rm{m}$; see Fig.~3).
Each lens belongs to an ommatidium\index{ommatitium},
which typically contains eight photoreceptors, and a
number of optical screening cells\index{photoreceptor}.
Part of the photoreceptor cell membrane is densely
folded, and forms a long ($\approx 100-200 \mu\rm{m}$),
thin ($\approx 2 \mu\rm{m}$) cylinder, called the
rhabdomere\index{rhabdomere}. The membrane consists
mainly of phospholipids and proteins so that its
refractive index is higher than that of the surrounding
watery medium. Therefore, the rhabdomere acts as a
waveguide, and light can travel along its long axis in a
bound mode. The combination of a lens and the tip of an
optical waveguide in its focal plane forms a spatial
filter (Snyder 1979): Only light coming from a small
angular region can enter the waveguide. The physical
limit to the resolution of this system is set by
diffraction: $\Delta \phi \approx \lambda /D$, and with
$\lambda=500$~nm and $D=25~\mu$m, we have $\Delta \phi
\approx 1/50~\rm{rad} \approx 1^{\circ}$. See Exner
(1891), Barlow (1952), and Feynman et al. (1963) for an
analysis of the optics of compound eyes. The physics of
the lens-photoreceptor system is well understood, and
theory is in very good agreement with physiological
findings (van Hateren 1984).

Flies do not have an iris pupil or eyelids,
yet they need to protect their photoreceptors from
excessively intense light. This is accomplished by an
elegant mechanism depicted in the top box in
Fig.~3. The top ommatidium shows
a dark adapted photoreceptor, which has tiny pigment
granules dispersed through its cell body.
In the light adapted state shown in the bottom
ommatidium these granules have migrated close to the
photoreceptor rhabdomere. The granules absorb light, and
because they are close to the light guiding rhabdomere,
they act as a light absorbing cladding that
captures up to 99$\%$ of the incoming photon flux. This
then prevents the photoreceptor from saturating in
bright daylight (Howard et al. 1987). The effectiveness
of the pupil \index{pupil} as a light absorber is
regulated by feedback (Kirschfeld and Franceschini 1969,
Franceschini and Kirschfeld 1976).

The photoreceptors in each ommatidium are numbered
R1-R8, and arranged such that R7 lies on the optical
axis of the lens. R8 lies behind R7, while the
rhabdomeres of R1-R6 lie in a trapezoidal pattern around
the center, as shown by the cross section in the top
frame in Fig.~3. The optical
axes of neighboring ommatidia point in slightly
different directions and they differ by an amount
that matches the angular difference among the
photoreceptors in a single ommatidium. Therefore,
eight photoreceptor cells in seven neighboring
ommatidia share a common direction of view. Receptors
R1-R6 in all ommatidia have the same spectral
sensitivity (Hardie 1985), and those R1-R6 receptors in
neighboring ommatidia that share the same direction of
view combine their signals in the next layer, the
lamina. This is known as neural superposition
\index{neural superposition} (Braitenberg 1967,
Kirschfeld 1967, 1979). Receptors R7/R8 are special,
having a spectral sensitivity different from R1-R6, and
bypassing the lamina, projecting directly to the
medulla.

\subsection{Reliability and adaptation}

\begin{figure}[tbp]
\begin{center}
\includegraphics[keepaspectratio,width=1.0\textwidth]
{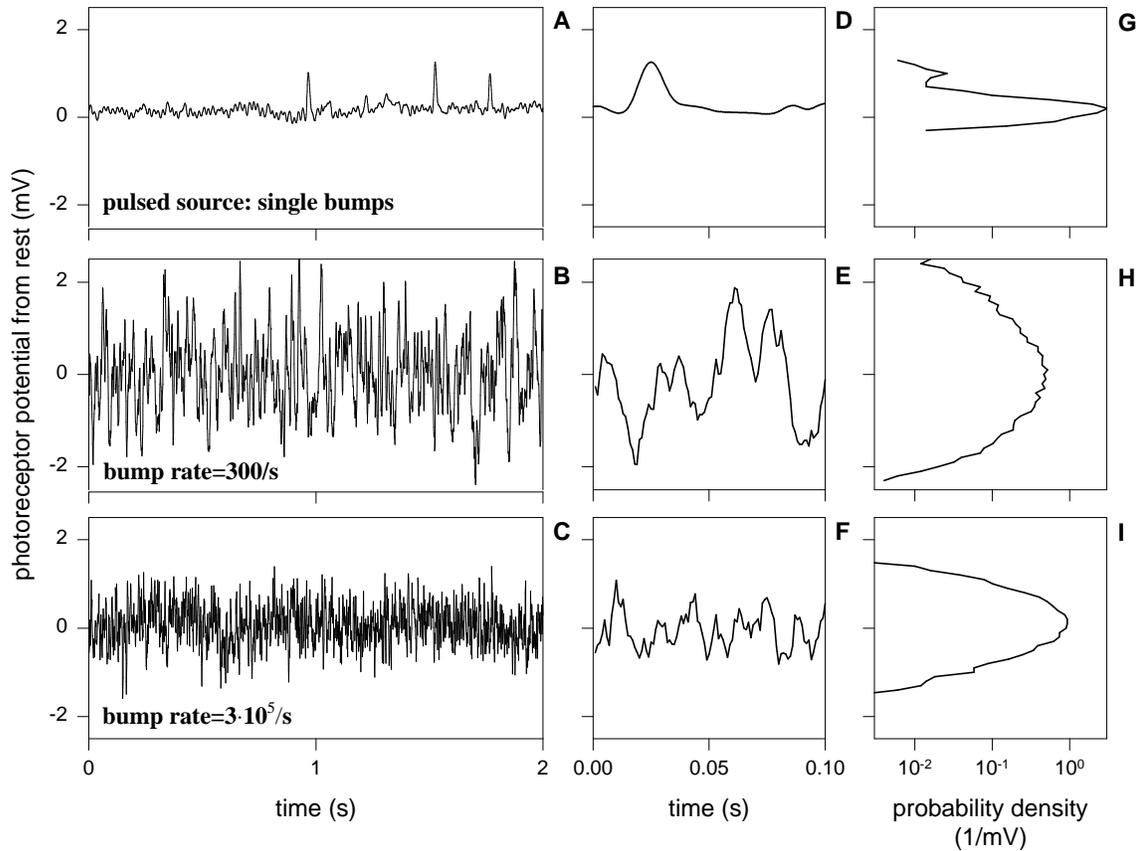}
\caption[ ]{\footnotesize Intracellular photoreceptor
  recordings. A,~B,~C: samples of
  photoreceptor voltage, each 2 seconds long. A:
  Recording in the dark adapted state, with a light source
  that emitted brief flashes leading to about 0.5 photons
  captured per flash on average. B: The same cell,
  with a continuous light source, and a photon capture
  rate of 300 per second. C: As B, but now at
  a 100 times higher light intensity. Panels D,~E,~F 
  show part of the traces in repectively A,~B,~C 
  but at higher time resolution. G,~H,~I: 
  Amplitude distributions of the
  signals in A,~B,~C. Note that these
  are drawn sideways, so that the vertical scales in all
  panels are the same. The probability densities are
  drawn on a logarithmic scale, which means that the
  curves should approximate parabolas for Gaussian
  amplitude distributions.}
\end{center}
\end{figure}

Because single photon responses are more or less
standardized it seems a good idea to model the
photoreceptor voltage as a shot noise process,
consisting of a stream of photon
induced events often referred to as ``quantum
bumps'' \index{quantum bump}\index{bump} (Dodge et al.
1968, Wong et al. 1980, de Ruyter van Steveninck and
Laughlin 1996a,b). However, we
are dealing with a highly adaptive system. As discussed
in the previous section, the variance of shot noise
should be proportional to the rate of the underlying
Poisson process. Panels B and C of Fig.~4 show that
the variance at a mean bump rate of 300/s is much higher
than at a rate of $3 \cdot 10^5/$s. So shot noise
\index{shot noise} does not appear to be a good model.
The solution to this dilemma is that the bumps change
shape so that both their amplitude and their width
decrease when the photoreceptor is exposed to higher
light intensity (see also Fig.~9). This is the gist of
the ``adapting bump model'' formulated by Wong et al.
(1980)\index{bump!adapting}. Clearly, when bumps adapt,
the shot noise model loses some of its generality in
terms of predicting the expected response amplitude to a
certain stimulus or of predicting the noise power
spectral density at different light intensities.
However, one much more crucial aspect remains---the 
signal to noise ratio. Even if bump amplitudes
adapt to the ambient light intensity, the
frequency dependent signal to noise ratio depends only
on the rate of the underlying Poisson process. One last
caveat is in place here: In the standard shot noise
model, the ``bump'' shape is taken to be fixed. Surely, in
any realistic system there is noise, and here one can
distinguish two of its net effects: A variation in the
amplitude of the bumps, which limits the reliability of
the system at all frequencies, and a loss in timing
precision, which affects the higher fequencies more than
the lower frequencies, as treated in
section~\ref{RANDFILT}. See Stieve and Bruns (1983) for
more details on bump shape and de Ruyter van Steveninck
and Laughlin (1996b) for its effects on the overall
reliability of the photoreceptor.

The results we present below were obtained
in experiments using a green light emitting
diode (LED) as a light source, which was always
modulated around the same mean light level. The mean
photon flux into the fly's photoreceptors was set by
using filters of known density in the
optical path between LED and fly. Light intensities were
calibrated by counting the rate of quantum bumps at a
low light intensity, as depicted in Fig.~4a . This
was generally done with the LED delivering
short (1 ms) flashes, using filters of optical density
between 3 and 4, that transmit a fraction of
$10^{-3}$ to $10^{-4}$ of the incident photons. From
this low level, light intensities were increased in
calibrated steps, and we can so define an extrapolated
bump \index{bump!extrapolated
rate}\index{rate!bump, extrapolated} rate for each
setting of light intensity in the experiments. Note that
this procedure does not specify the absolute quantum efficiency
of photon detection. At low light levels this
efficiency, defined as the
probability for an on axis photon of optimal wavelength
(490~nm) to lead to a quantum bump, is about 50\% (Dubs
et al. 1981, de Ruyter van Steveninck 1986).

In most experiments the LED was
modulated with a pseudorandom
waveform  of
duration 2s, sampled at 1024~Hz with independent
values at all sample times. This waveform was
constructed in the frequency domain, by computing a
series of 1024 random numbers on $[0,2\pi)$,
representing the phases of Fourier components. That
list, concatenated with an appropriate list of negative
frequency phase components, was inverse Fourier
transformed, resulting in a sequence of real numbers
representing a series of 2048 time samples that was used
as the contrast signal $c(t)$. This procedure ensured that the
amplitudes of all frequency components were equal. This
is not necessary in principle, but it is very convenient
in practice to have a signal that is free from spectral
dropouts.
\index{contrast!Gaussian modulation}
The use of pseudorandom waveforms to study neural signal
processing has a long history; for a review see Rieke et al. (1997).
In the photoreceptors and LMCs of the fly, the first such experiments
were done by French and J\"{a}hrvilehto (1978).

\begin{figure}[tbp]
\begin{center}
\includegraphics[keepaspectratio,width=1.0\textwidth]
{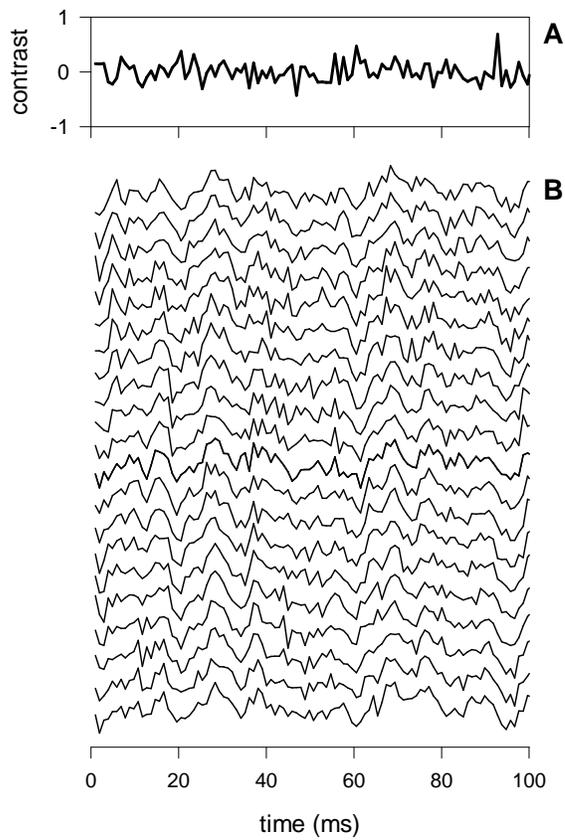}
\caption[ ]{\footnotesize 
  A: Modulation of the LED light source by
  a computer generated pseudorandom signal sampled at
  $\approx$1 ms intervals. This panel shows a small,
  100~ms section of the total trace which was 2~s long.
  In the experiment the contrast waveform depicted here is
  presented repeatedly, and the response of the cell to
  all these repetitions is recorded. B: Twenty
  examples of individual traces recorded from a blowfly
  photoreceptor in response to the contrast
  waveform shown in A.}
\end{center}
\end{figure}

Fig.~5 summarizes the measurements made in a typical
experiment on a photoreceptor. The same contrast
waveform (top) was presented repeatedly and the response
to each presentation recorded. From a large number of
repetitions we obtain an ensemble of responses, from
which we compute the ensemble average \index{ensemble
average} shown in Fig.~6A. Clearly the traces in Fig.~5B
share the same overall shape, but are not identical.
Each trace differs from the ensemble average, and these
fluctuations \index{fluctuations} represent the noise in
the system. An ensemble of noise traces is obtained simply
by subtracting the ensemble averaged response
from each individual trace. 
\begin{figure}[tbp]
\begin{center}
\includegraphics[keepaspectratio,width=1.0\textwidth]
{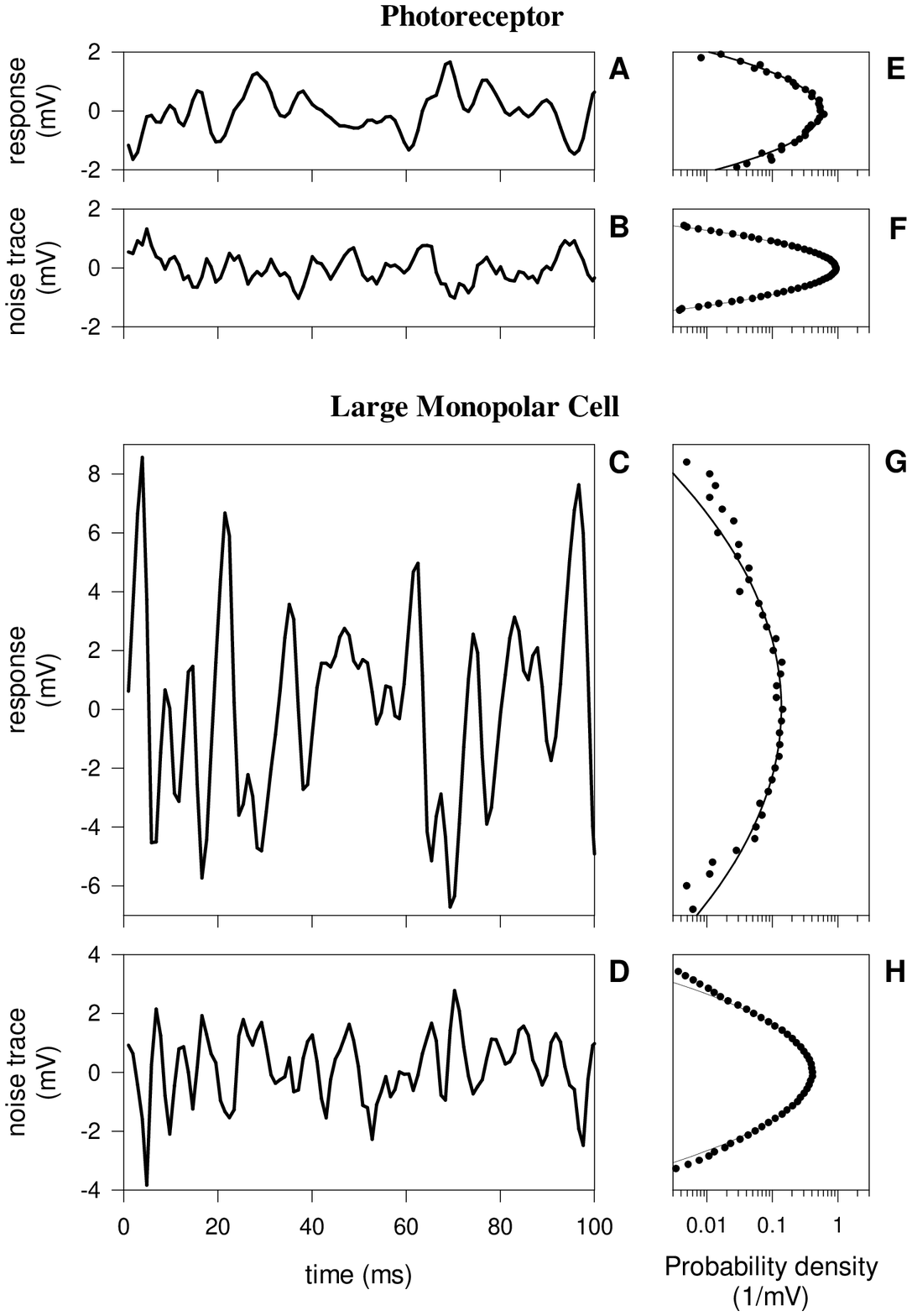}
\caption[ ]{\footnotesize 
  Sample traces of an experiment on a
  photoreceptor and an LMC, and amplitude
  distributions. A: A 100 ms segment of the
  ensemble averaged response of a photoreceptor
  to the modulation waveform shown in
  Fig.~5A. B: Example of a 100 ms
  noise trace, that is the difference
  between one trace in Fig.~5B
  and trace A in this figure. C,~D As
  A,~B  above, but for an LMC. Note that
  the vertical scales in panels A-D  are all
  the same: The LMC's signal is much larger than that of
  the photoreceptor. E: Amplitude distribution
  of the average voltage trace of the
  photoreceptor. Dots: measured values; line: Gaussian
  fit. F: As E, but now for the photoreceptor
  noise waveform. G,~H: as E,~F but for the LMC.}
\end{center}
\end{figure}

\index{contrast!transfer function}

We first characterize the cell's linear response
by computing
the ratio of the Fourier transform of the
ensemble averaged voltage waveform $V_h(t)$, and the
Fourier transform of the contrast modulation waveform
$c(t)$, as in Eqn.~\ref{eq:CTRF}. The square of the
absolute value of this ratio gives us the contrast power
transfer function.
The ensemble of noise traces is described by a power
spectral density,\index{noise power} which we
find by averaging the Fourier power spectra of all the
individual noise traces in the experimental ensemble, 
or equivalently by computing the variance of each Fourier coefficient
across the ensemble of noise traces. Finally, the ratio of
these two functions, as defined in Eqn.~\ref{eq:RATE_FLUCT_EST}, 
is the effective Poisson rate.\index{Poisson rate!effective}
If the photoreceptor
were an ideal photon counter, the ratio should not
depend on frequency, and be numerically equal to the
extrapolated bump rate. Fig.~7 shows that the ratio
does depend on frequency, and that it goes down at the
higher frequencies, notably so above 100 Hz. This is a
consequence of the limited temporal resolution of the
transduction cascade which after all consists of a
sequence of chemical reactions, each of which depends on
random collisions between molecules. However, at
$3.8\cdot 10^3$ incoming bumps per second the
photoreceptor acts essentially as an ideal photon
counter up to about 50~Hz. The deviation from this at
low frequencies is due to excess low frequency noise in
the noise power density (see Fig.~7B,E) which is almost
certainly due to $1/f$ noise in the equipment. At the
higher extrapolated bump rate the photoreceptor's
efficiency is constant up to almost 100 Hz, and is about
a factor of 2 from ideal. The behavior of the LMC is a
bit different in that it acts as a high pass filter, and
transmits low frequencies less reliably than would an ideal
detector. However, for intermediate frequencies
the LMC stimulated at a bump rate of $7.5\cdot10^4$ per
second comes close to ideal. At the highest bump rate,
$7.5\cdot10^6$ per second, the LMC deviates from ideal 
by a factor of 8 at its best frequency ($\approx 50$ 
Hz). At this frequency the absolute performance of the 
LMC is quite impressive, being equivalent to an 
effective photon flux of about $10^6$ events per second.

\begin{figure}[tbp]
\begin{center}
\includegraphics[keepaspectratio,width=1.0\textwidth]
{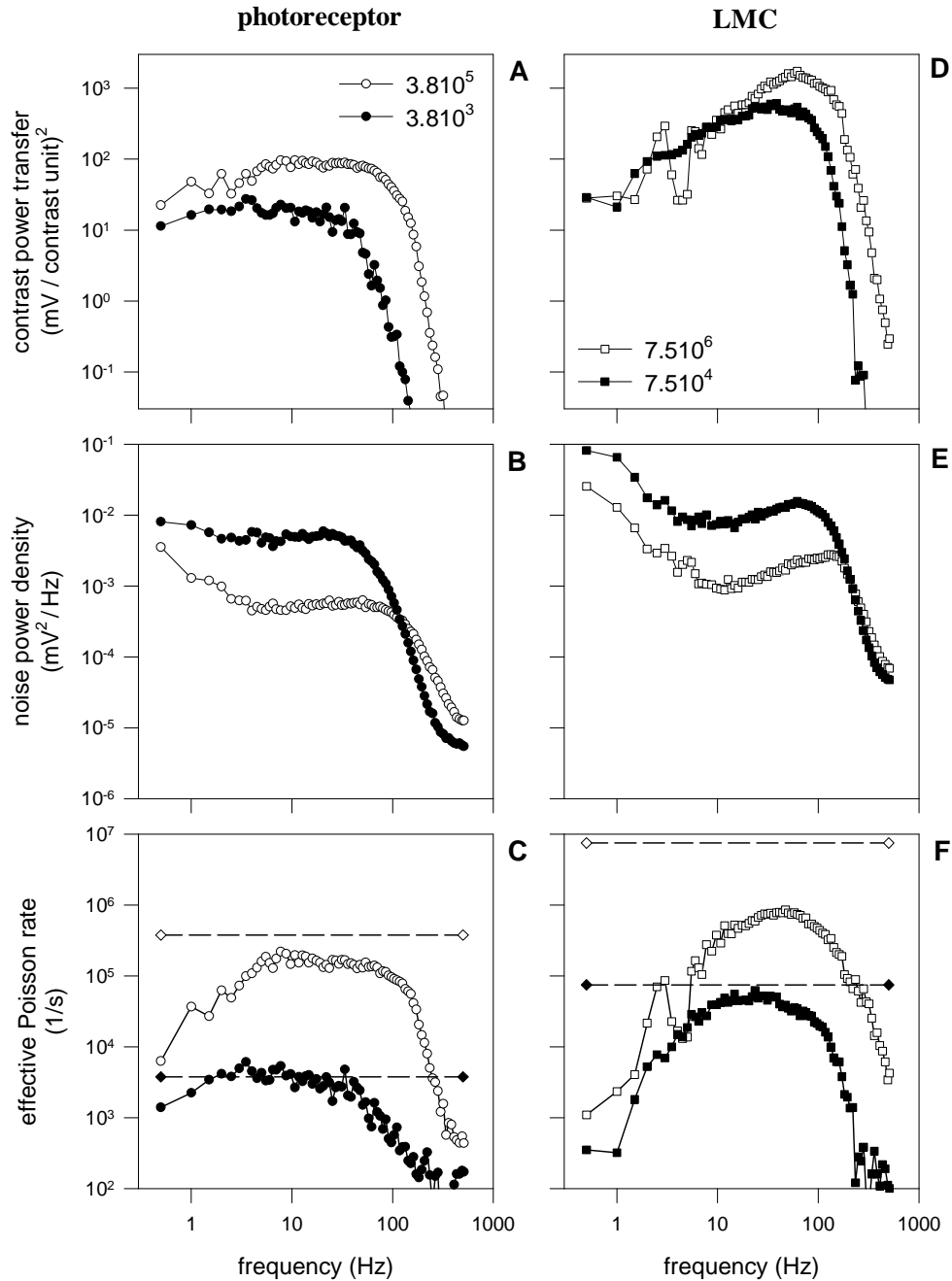}
\caption[ ]{\footnotesize Characterization
  of a photoreceptor and an LMC in the frequency domain,
  each for two light intensities. The legends of A
  and B give these intensities, expressed as
  extrapolated bump count rates. A: Photoreceptor
  contrast power gain as a function of frequency. B:
  Power spectral densities of the ensemble of noise traces
  represented by Fig.~6B. C: The effective
  Poisson rate, calculated as the ratio of contrast power
  gain (panel A) and the power spectral density
  (panel B) at each frequency. If the
  photoreceptor were an ideal photon counter, not adding
  any noise itself, then the effective Poisson rate should
  be spectrally flat, at a level of the extrapolated bump
  rate given by the legend in panel A. This is
  depicted by the dashed lines in C. D,~E,~F: as 
  A,~B,~C, but now for an LMC.} 
\end{center} \end{figure}
\begin{figure}[tbp]
\begin{center}
\includegraphics[keepaspectratio,width=1.0\textwidth]
{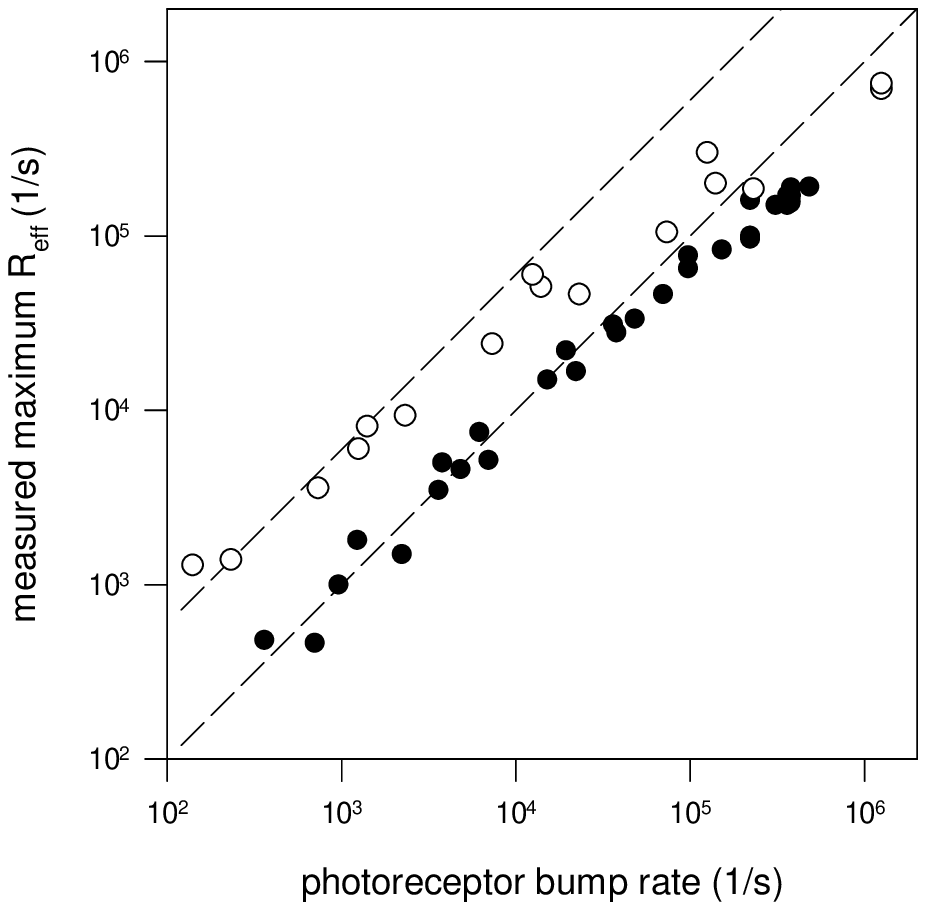}
\caption[ ]{\footnotesize Comparison of the best measured
  statistical performance of photoreceptors and LMCs to
  the theoretical limit imposed by photon shot noise. The
  peak values of the effective Poisson rate curves, such
  as those in Fig.~7C,F are plotted as a function of
  the rate at which photons are absorbed by a
  photoreceptor. The rate of photon absorption is
  extrapolated from a dark adapted experiment, in which
  individual bumps are counted (see Fig. 4A). The dashed
  lines represent the behavior of an ideal photon counter
  for photorecptors (lower line) and LMCs (upper line).
  Along the abscissa the bump count calibration values of
  the LMC are divided by 6 to get the bump rate of each of
  the presynaptic photoreceptors. Statistically efficient
  neural superposition requires that the LMC uses all
  photons from all six photoreceptors from which it
  receives input, which means that the LMC data points
  should then follow the upper dashed line.}
\end{center}
\end{figure}

Fig.~8 presents an overview of the cell's best
performance, and compares it to the theoretical limit.
Here we show the maximum of the effective Poisson rate
for six photoreceptors and three LMCs, each at multiple
light levels. These data are plotted as a function of
the photoreceptor's extrapolated bump rate, which for
LMCs is the extrapolated LMC bump rate divided by six,
because it receives input from six photoreceptors. For
the photoreceptors, at low light levels, the measured
maximum and the extrapolated bump rate are very close,
and they are at most a factor of two apart at the
highest light intensities measured here. This indicates
that the photoreceptor is designed to take advantage of
each transduced photon, up to fluxes of about
$3\cdot10^5$ extrapolated bumps/s, and up to about 100 Hz.
At bump rates around $10^5$/s and higher the efficiency
of the photoreceptor begins to decline.
\index{photoreceptor!efficiency}
\index{efficiency!photoreceptor and LMC} This is caused
primarily by the action of the pupil
\index{pupil} (Howard et al. 1987; see also
section~\ref{Optics}), which attenuates the photon flux
propagated through the rhabdomere and thus increases the
effective contrast noise as described in
section~\ref{RANDFILT}.

One reasonably expects that if the fly's brain is well
designed, it would put the relatively high quality of the
photoreceptor signals to good use. This would then imply
that the accuracy of visual information processing is
not too far from the photon shot noise limit at the
light levels studied here. A first check is to see if
neural superposition \index{neural superposition} is
efficient in this statistical sense.   Comparing the
measurements of maximal LMC effective photon flux with
the ideal (open symbols and top dashed line in Fig.~8),
we see that this is indeed the case up to photoreceptor
bump rates around $10^4$--$10^5$ per second. For higher
light intensities the LMC, as the photoreceptor, becomes
less efficient. There is a hint in the data that the
LMC declines somewhat faster than the photoreceptor, 
perhaps due to limitations in the reliablity of synaptic
transmission, as was noticed by Laughlin et al. (1987).

Given that neurons have a dynamic range
much smaller than typical sensory signals, the
question arises of how the nervous system copes with the
input it receives in order to transmit and process
information efficiently. Ultimately this is a
matter of optimal statistical estimation, and
the result will therefore depend on the statistical and
dynamical characteristics of the signals that the animal
encounters in its environment. It is well known, and we
have seen examples already, that photoreceptors adapt
their gain to the ambient light intensity, which may
vary over many orders of magnitude. The usefulness of
this adaptation \index{adaptation} seems obvious: At
higher light intensity the absolute gain should be
brought down to keep the transduced contrast
fluctuations within the cell's voltage operating range.
But implicit in this explanation is that the average
light levels must change relatively slowly compared to
the contrast fluctuations. If the sun flickered
unpredictably on time scales of a second or so, and with
an amplitude equal to that of the day night cycle, then
it would be of no use at all to adapt the photoreceptor
gain. Only because large changes tend to be slow does it
make sense to design a system that tracks their mean,
and changes its gain so as to encode the faster
fluctuations more efficiently.

\begin{figure}[tbp]
\begin{center}
\includegraphics[keepaspectratio,width=1.0\textwidth]
{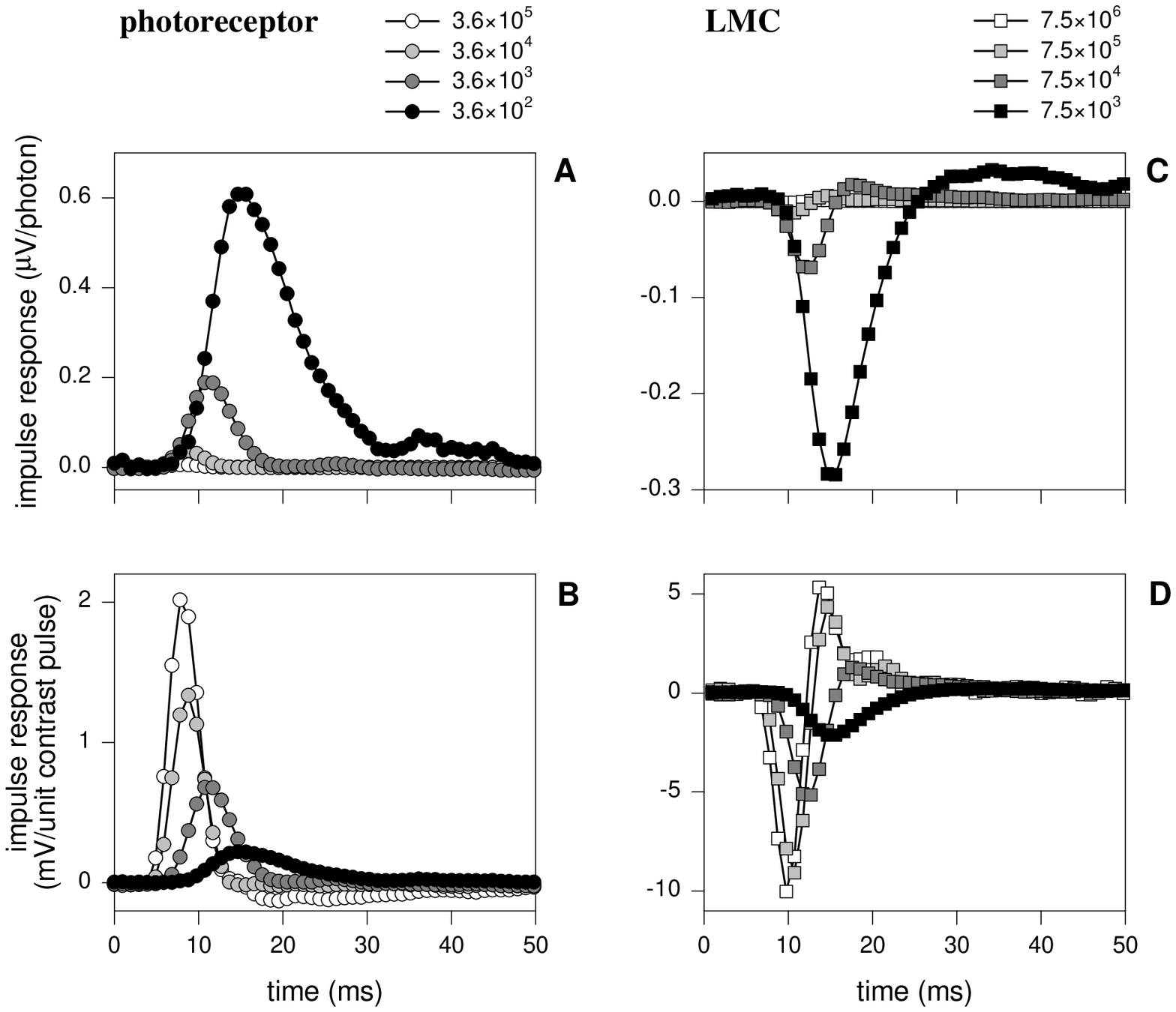}
\caption[ ]{\footnotesize 
  Impulse responses of blowfly photoreceptors and LMCs,
  measured in the adapted state at different light
  intensities. The average light intensity, expressed as
  the extrapolated rate of effective photon capture
  in bumps/s, is given in the legends above A and
  C. The responses are scaled in two different ways.
  A: Photoreceptor impulse response, scaled to
  represent the electrical response of a photoreceptor to
  the capture of a single photon. B: The same data
  as in A, but here expressed as the electrical
  response to a contrast pulse of 1 ms wide, with an
  amplitude equal to the mean light intensity. C,~D:
  as A,~B, but for an LMC.}
\end{center}
\end{figure}

The effect of adaptation is easier to appreciate when
the responses are
plotted in the time domain. Fig.~9 shows
impulse responses \index{impulse response} of both cell
types, at four light levels spanning three orders of
magnitude. All the curves are derived from inverse
Fourier transforms of the contrast transfer function
$\hat{H}(f)$ defined earlier. The top two panels show
the absolute gain---that is the response normalized to a
single photon capture. In the bottom panels the response
is normalized to a 1 ms lightflash with an amplitude
equal to the mean light level, so these are normalized
to contrast. It is clear from the figures that there is
a large range of gain control, as the shapes of the
responses normalized to single photons vary dramatically.
Expressed as contrast gain, the responses become higher
in amplitude and sharper in time as the light intensity
increases.

The photoreceptor primarily seems to scale down its
photon conversion gain, both in amplitude and in time
course. The LMC filters out the photoreceptor's DC
level, except perhaps at the lowest light intensities,
and it also scales the gain (Laughlin and Hardie 1978).
The combined effect leads to a scaling of the LMC
amplitude fluctuations. Remarkably, this scaling is such
that when stimulated with naturalistic, and very
non-Gaussian, sequences, the LMC produces a voltage
output that follows a Gaussian quite closely (van
Hateren 1997). There is  evidence that
adaptation of this system is set so as to optimize
information transmission rates under different
conditions (van Hateren 1992).

The data of Fig.~9 show the behavior of the cells while
they are in their adapted states, as care was taken to
to let the system adapt before the measurement was done.
It is also interesting to study the time course of
adaptation, and here we will look in particular at
adaptation of synaptic transmission.

\subsection{Efficiency and adaptation of the
photoreceptor-LMC synapse} \label{PRLMCSYN}

The link between the photoreceptors and the LMC is a
parallel array of chemical synapses.
\index{synapse!blowfly photoreceptor-LMC} The detailed
anatomy of this projection is well known (reviewed by
Shaw 1981), and counts have been made of the number of
active zones between photoreceptors and LMCs. The total
number of synapses between one photoreceptor and an LMC
is on average about 220 (Nicol and Meinertzhagen 1982,
Meinertzhagen and Fr\"{o}hlich 1983), so that the total
number of active zones feeding into an LMC is close to
1320. Although this number was obtained from the
housefly \emph{Musca}, which is smaller than the blowfly,
the total number is unlikely to be far off. The
photoreceptor-LMC synapses are tonically active, just
like the synapses of vertebrate retinal bipolar cells.
This means that even in the absence of contrast
fluctuations, they release a stream of vesicles
\index{synapse!tonically active} (Shaw 1981,
Uusitalo et al. 1995, Lagnado et al. 1999). Also they
pass on graded potentials, and it thus seems a
reasonable first approximation to model them as units
that release vesicles with a rate depending on the
presynaptic potential. One interesting question is then
whether vesicle release can be modeled as a modulated
Poisson process.\index{vesicle release!and
modulated Poisson process} At first sight this would
seem wasteful, in the sense that if the synapse would
have better control over its vesicle release, it
could emit vesicles in a much more deterministic
way. One might imagine that the synapse functions
somewhat as a voltage controlled oscillator, releasing
vesicles in a regular stream at a rate determined by the
presynaptic voltage. One way to get at this issue is to
measure the reliability of the synapse, and estimate
from this a lower bound on the release rate, assuming
that release is Poisson. If, through other independent
methods, one can make a good estimate of the average
total release rate, then one can compare the two rates.
If the rate estimated from the Poisson assumption were
found to be much higher than the total average rate, one
would have a strong indication for tight control of
vesicle release.

As argued in section~\ref{CTRF_ECN}, one can find the
equivalent contrast noise power of a cascaded system by
adding the equivalent contrast noise power of its
separate elements $N_c(f)=N_{c1}(f)+N_{c2}(f)$. What we
would like to do here is to infer the equivalent
contrast noise power for the synapse from measurements
of the equivalent contrast noise of the photoreceptor
and the LMC. We have the data, because the equivalent
contrast noise power is just the inverse of the
effective Poisson rate plotted in
Fig.~7C,F: $N_c(f)=1/\lambda_{\mathrm{eff}}(f)$ (see
section~\ref{CTRF_ECN}).
The measurements of effective Poisson rates in
photoreceptors and LMCs, together with the given number
of active zones, allow us to make an estimate of the
effective Poisson rate of a single synaptic contact.
\index{Poisson rate!effective for synapse}
This cannot be done directly because it has not been possible
in practice to make a simultaneous recording from a photoreceptor and
its
postsynaptic LMC in vivo, so we have to interpolate. From a large number
of experiments on different cells at different light levels we compute
$N_c(f)$, and we do this separately for photoreceptors
and LMCs. We interpolate each set of curves to
find a smooth surface describing the overall behavior as
a function of both frequency and bump rate (see de
Ruyter van Steveninck and Laughlin 1996a). We estimate
the synaptic contribution, $N_{c_{\mathrm{Syn}}}(f)$ by
subtracting the interpolated values obtained for
photoreceptors, divided by 6 to account for the parallel
projection of 6 photoreceptors, from those describing
LMCs:
$N_{c_{\mathrm{Syn}}}(f)=N_{c_{\mathrm{LMC}}}(f)-N_{c_{\mathrm{PR}}}
(f)/
6$.
The differences are small, and not so easy to estimate,
which already indicates that the
synaptic array itself cannot be much less
reliable than the photoreceptor. Here we only present
data from the highest light levels used in the
experiment, mainly because there all signals are most
reliable, and the effect of internal noise sources is
most conspicuous.

Note that $N_{c_{\mathrm{Syn}}}(f)$ describes the
equivalent contrast noise of the full array of 1320
synapses in parallel. Each synapse is driven by the same
photoreceptor voltage fluctuations, and each modulates
its vesicle release rate accordingly. We assume
that apart from this common driving force, all synapses
release vesicles in a statistically independent way, and
that all are equally effective and reliable. Then we
can simply divide $1/N_{c_{\mathrm{Syn}}}$, as defined
above, by 1320 to get the effective Poisson rate for the
single synapse, as argued in Sect.~\ref{CTRF_ECN}. This
then provides a lower bound on the reliability of
a single synapse. If the assumptions mentioned here are
invalid, then there must be at least one synapse in the
array that does better than this average.

\begin{figure}[tbp]
\begin{center}
\includegraphics[keepaspectratio,width=1.0\textwidth]
{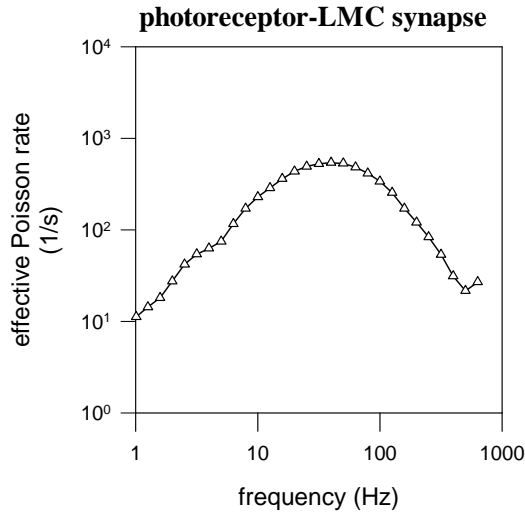}
\caption[ ]{\footnotesize 
  Effective Poisson rate for a single synaptic
  active zone as a function of frequency, computed as
  described in the text.}
\end{center}
\end{figure}

Fig.~10 shows the result of the
calculation, expressed as $1/N_c(f)$, the equivalent
Poisson rate for one single active zone. The curve has a
maximum of about 540 events per second per synaptic
zone. Unfortunately we cannot directly identify this
number with the supposed vesicle release rate.
This can be done for photon flux
modulations because in that case we know the proportion, $c(t)$,
by which we modulate the flux: $\lambda(t)=[1+c(t)]\cdot
\lambda_0$. For the photoreceptor-LMC synapse, one may
describe the modulation of the
release rate by a gain
factor $g$ (which may also be frequency dependent) that
converts photoreceptor voltage into vesicle flux
(Laughlin et al. 1987). If $g$ is high, then the synapse
encodes relatively reliably at a low mean vesicle rate.
The price is that the operating range will be small, as
the rate cannot go below zero. If $g$ is low, on
the other hand, then the operating range is large, but
the reliability of transmission is relatively poor. In
other words, we measure an effective rate
$\lambda_{\mathrm{eff}}$, from which we wish
to estimate a physical Poisson event rate. This means we
must get an estimate of $g$, and we must understand how
$g$ affects our measurement. To begin with the latter,
if we just apply Eqn.~\ref{eq:RATE_EST} to the case of a
real rate $\lambda_0$ modulated by $g\cdot c$, then we
would measure an effective rate $\lambda_g$ depending on
$g$:

\begin{equation}
\label{eq:GDEP}
\lambda_g = \lambda_0 \cdot |g|^2,
\end{equation}

\noindent while release would shut down at contrast
values below $c_{\mathrm{min}} = -1/g$. To keep the
argument simple we neglect the possible frequency
dependence of $g$. That is justified here because
we will not reach precise
conclusions anyway, and the frequency dependence, being
rather smooth, is not likely to affect the
final result of the analysis too much. We
need to find a way to estimate the operating range of
the synapse, and we can try to get at that by estimating
$c_{\mathrm{min}}$. One hint is that LMCs are reasonably
linear when the contrast fluctuations are not too large,
perhaps of order 20\% to 30\%, but we would like to make
this a bit more precise.

\begin{figure}[tbp]
\begin{center}
\includegraphics[keepaspectratio,width=1.0\textwidth]
{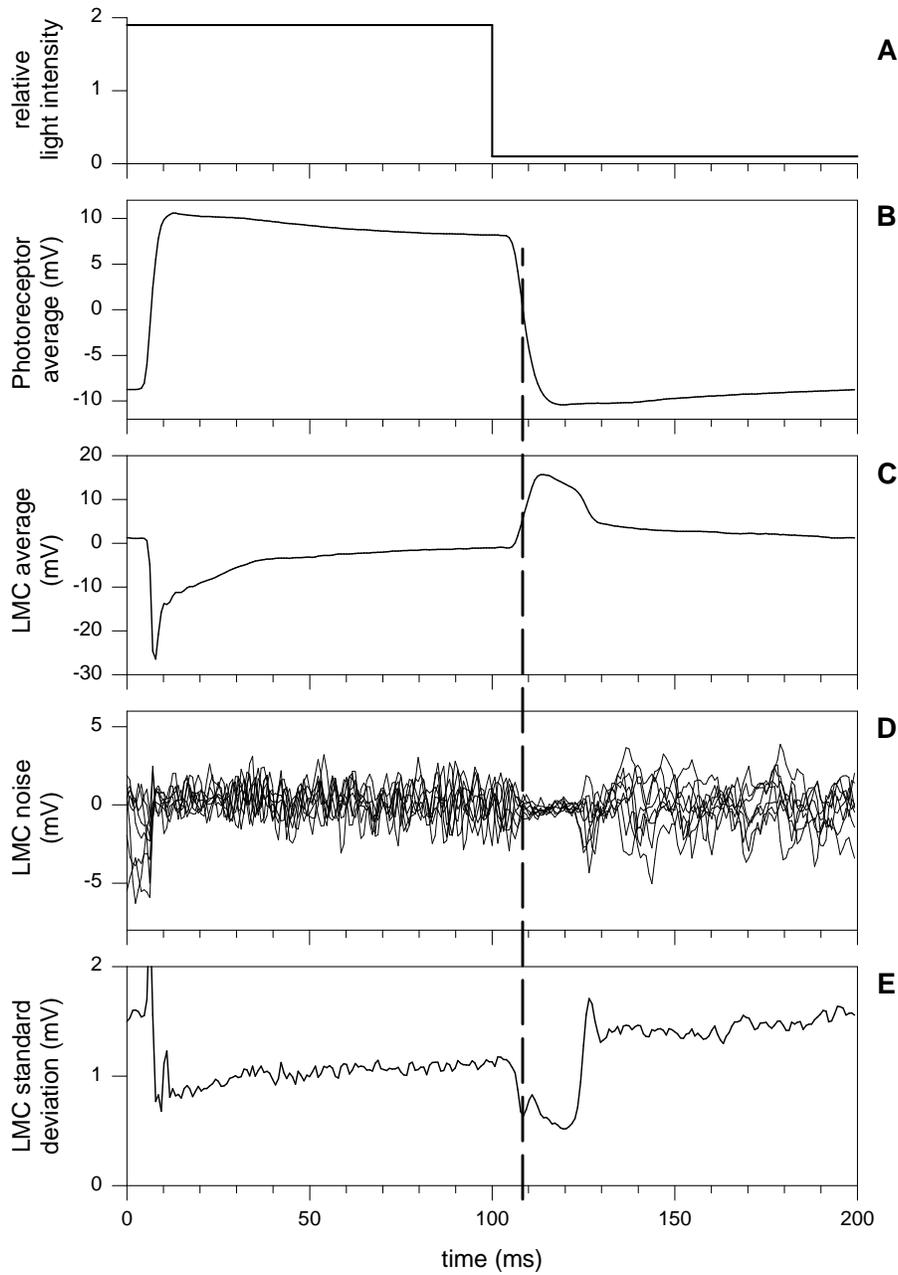}
\caption[ ]{\footnotesize 
  Averaged responses of a photoreceptor and an
  LMC, and LMC fluctuations in response to a large
  amplitude modulation. A: The stimulus
  contrast sequence is a square wave of amplitude 0.95,
  and duration 200~ms, sampled in 256 bins at 1280~Hz.
  This waveform is presented 380 times, while the response
  of a photoreceptor or an LMC is recorded. B:
  Ensemble averaged photoreceptor response. C:
  Ensemble averaged LMC response. D: Example of 8
  traces showing fluctuations of the LMC response around
  its average waveform. E: Time dependent standard
  deviation of the LMC fluctuation traces. The standard
  deviation plotted here at each instant of time is the
  standard deviation across the ensemble of LMC voltage
  fluctuations, all taken at the same phase of the square
  wave stimulus.}
\end{center}
\end{figure}

Fig.~11 presents data
suggesting that we can see the synapse shutting
down\index{synapse!shutdown of transmission}. The
stimulus (Fig.~11A) is a 200 ms square wave of 95\%
contrast, repeated 360 times. Panel B shows the average
photoreceptor response which follows that stimulus with
a bit of sag. The LMC (panel C) responds phasically and
with inverted sign. When the photoreceptor voltage makes
its downward transition, the vesicle release rate
decreases, and because the neurotransmitter (histamine,
Hardie 1988) opens chloride channels the
LMC depolarizes. In panel D we plot the fluctuations (8
samples) of the LMC potential around its average
waveform. These show  a rather dramatic effect just
after the light to dark transition. It seems that
synaptic transmission is completely shut down when the
photoreceptor hyperpolarizes, and bounces back about 15
ms later, very similar to results reported by Uusitalo
et al. (1995). This is confirmed by the standard
deviation of the fluctuation waveforms shown in panel E.
The fluctuations during constant light, say from 20-100
ms and from 130-200 ms, are due to a combination of
photoreceptor noise amplified by the synapse and
intrinsic noise of the vesicle release itself. We can
also add a little probe signal to the large square wave
stimulus, and we see a similar effect in the gain of the
synapse: The photoreceptor fluctuations are not
transmitted during the same 15 ms window, as can be
seen in Fig. 12. The apparent shutdown coincides
with the photoreceptor voltage being halfway between the
light and dark adapted value, which in turn is induced
by an intensity drop of almost 100\%. Shutdown thus
seems to correspond to about 50\% modulation, in other
words, $g\approx2$. Combined with our earlier estimate
of an effective rate of 540/s, from Eqn.~\ref{eq:GDEP} we
get a rate of $540/2^2=135$ vesicles per second, based
on the measured reliability. This is lower than the rate
of 240 per second reported by Laughlin et al. (1987),
but in view of the errors the discrepancy is not too
surprising. We would like to compare the more
conservative estimate of this number with a more direct
measurement of average release rate. Unfortunately there
are no conclusive data on synaptic release rates of the
photoreceptor-LMC synapse in the fly.
\index{synapse!vesicle release rate} There are
experimental estimates of tonic release rates for
goldfish retinal bipolar cells which, like the fly
photoreceptor, transmit graded potentials across a
chemical synapse. In a recent paper, Lagnado et al.
(1999) report 23 vesicles/s in this system, a factor of
6 lower than what we estimate here.

The numbers we derive here are certainly not 
precise, and a comparison between very different species
is always tenuous. Although no hard conclusions
can be drawn, the comparison points to an interesting
possibility. The discrepancy between the high Poisson
release rate required to explain the reliability on the
one hand, and the lower measured release rates in
goldfish on the other, is large. The most interesting
explanation for this, in our view, is that the Poisson
assumption is not valid, and that the synapse would be
capable of releasing vesicles with much higher precision
than expected from that.

Of course Fig.~11 points out another
interesting aspect of transmission by the synapse,
namely that it is highly
adaptive.\index{synapse!adaptation} The
synapse seems to reset itself to follow the large swings
in the DC component of the photoreceptor voltage,
presumably to be able to encode fluctuations around this
average more efficiently. It has been known for a long
time that photoreceptor-LMC transfer is adaptive
(Laughlin and Hardie 1978, Laughlin et al. 1987), and
this has been interpreted as a resetting of parameters
to optimize information transmission (van Hateren 1992,
Laughlin 1989). Here we take a look at how fast this
type of adaptation takes place.

\begin{figure}[tbp]
\begin{center}
\includegraphics[keepaspectratio,width=1.0\textwidth]
{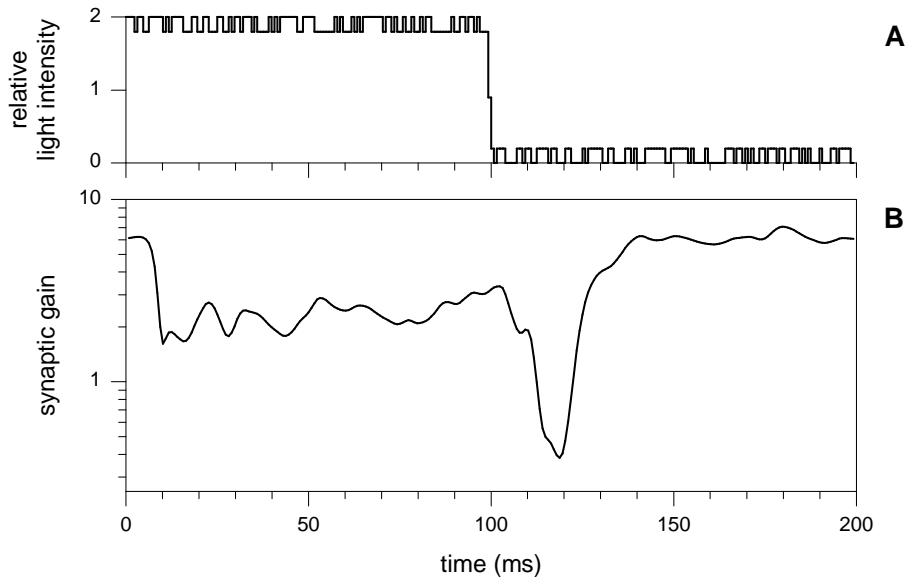}
\caption[ ]{\footnotesize 
  Characterization of the time dependent synaptic
  gain when the fly is stimulated with a 200 ms square
  wave pattern. A: Example of a combined waveform.
  To measure the gain of the photoreceptor and LMC
  along the 200 ms stimulus of
  Fig.~11, a small binary probe signal
  was added to the square wave. An example of the combined
  contrast waveform is shown here. B: Gain of
  synaptic transmission between photoreceptor and LMC, as
  the system cycles through the 200 ms period. See text
  for further explanation.}
\end{center}
\end{figure}

The experiment consists of the same square wave contrast
modulation as shown in Fig.~11, but
now a small amplitude random binary probe is added to
the large waveform (see Fig.~12). At
each presentation of the 200~ms square wave the probe
is a series of 256 binary random values, played
out at a sampling rate of 1280~Hz, and its
random sequence is different at each presentation. The
responses of the cell are recorded, also at 1280~Hz
sampling rate. After the experiments we correlate the
probe stimulus with the voltage variations
$\Delta v_m(t)=v_m(t)-V(t)$ it induces in the response, 
at consecutive points along the 200 ms square wave. That
is, we compute:

\begin{equation}
\label{eq:PRXCORR}
\Phi_{p\Delta v}(t,t')=
\lim_{M\to\infty}\frac{1}{M}\sum_{m=1}^M p_m(t)
\times \Delta v_m(t') \end{equation}

\noindent where the $m$ stands for
the different presentations, and $\times$ represents an
outer product. Thus $\Phi_{p\Delta v}(t,t')$ is the
crosscorrelation between probe stimulus $p(t)$ taken at
time $t$, and voltage response to the probe, 
$\Delta v(t')$, taken at $t'$, which we find by 
computing the ensemble averaged outer product of the 
probe with the response. In practice we treat $p_m(t)$ 
and $\Delta v_m(t)$ as vectors of 256 elements each, 
spanning the 200~ms repeat period, and 
$\Phi_{p\Delta v}(t,t')$ is then a 256$\times$256 
crosscorrelation matrix. We make the same measurement in
a photoreceptor and an LMC, and so get two 
crosscorrelation matrices. As before, we regard the 
photoreceptor and the LMC as a cascaded system, and, as 
long as things are linear, one should be able to 
describe the cascade as the following matrix 
multiplication:

\begin{equation}
\label{eq:CASCADE}
\Phi_{p\Delta 
v_{\mathrm{LMC}}}=\Phi_{p\Delta v_{\mathrm{PR}}} 
\cdot \Phi_{p\Delta v_{\mathrm{Syn}}}. 
\end{equation}

\noindent Note that we do not suggest that the system is
linear in the response to the large amplitude square
wave. It definitely is not. But what we try to
characterize here are the small fluctuations
due to the probe around the large amplitude average
waveform induced by the square wave. That can be
reasonably assumed linear, but nonstationary as a result
of the square wave.\index{nonstationarity!of synaptic
transmission}\index{linearity!and nonstationarity} This
nonstationarity then naturally leads to two time
indices, and they are both represented in the matrix
formulation. Ideally, from Eqn.~\ref{eq:CASCADE} we
should be able to derive the synaptic cross correlation
matrix by inverting the photoreceptor matrix, and
multiplying it with the LMC matrix. Unfortunately, this
procedure is rather unstable, both because of experimental
noise, and because of the strong high frequency components
in the LMC signal. Therefore we will settle for
something more modest here, and calculate the probe
induced variance. The stimulus induced time dependent
variance is computed as the diagonal of the
probe--response crosscorrelation matrix multiplied by its
transpose:

\begin{equation}
\label{eq:STINDVAR}
\sigma_{p\Delta v}^2(t)=\left[\Phi_{p\Delta v}(t,t')
\cdot\Phi_{p\Delta v}(t,t')^T\right]
\cdot \delta(t-t'),
\end{equation}

\noindent
and to compute the synaptic contribution we now
take the ratio of the LMC and the photoreceptor
diagonals, or if we want to express the linear gain
we take the square root of this quantity:

\begin{equation}
\label{eq:SYNGAINSQ}
\sigma_{p\Delta v_{\mathrm{Syn}}}(t)=
\frac{\sigma_{p\Delta v_{\mathrm{LMC}}}(t)}
{\sigma_{p\Delta v_{\mathrm{PR}}}(t)},
\end{equation}

\noindent
as shown in Fig.~12B. The figure again suggests
a shutdown of synaptic transmission during the falling
phase of the photoreceptor voltage. Further, the overall
gain as defined here switches from about 2.5 during the
bright phase to about 7 during the dim phase. The data
indicate that this switch in gain is also accompanied by
a change in shape of the synaptic impulse response,
which seems to become sharper and more biphasic during
the bright phase (see Juusola et al. 1995). To some
extent these effects are also seen in the photoreceptor
response, whose gain decreases and speeds up. In the
photoreceptor that is presumably due in large part to
the change in membrane conductance accompanying the
change in membrane potential. In the case of the synapse
it seems likely that the dynamics of vesicle
release changes, and this interpretation is supported
by the apparent shutdown in transmission.

\section{Coding in a Blowfly Motion Sensitive Neuron}

Thus far we have considered the reliability and
precision of phototransduction and synaptic
transmission, two of the first steps in vision. Now
we want to look ``deeper'' into the brain, to a point
where some nontrivial computations have been done. There
are two very different questions. First, we are
interested in the precision of the computation itself:
Is the brain making use of all the information available
in the array of photoreceptor signals? Second, we want 
to understand the way in which the brain represents the
results of its computations: What is the structure of
the ``neural code''? For an accessible example of these
issues we turn to the visual motion sensitive neurons in
the fly's lobula plate. 

The fly's lobula plate contains a number of motion
sensitive cells, shown in Fig.~3, that are direction
selective and have wide visual fields. They are thought
to achieve wide field sensitivity by adding the
contributions of a large number of small field motion
sensitive cells from the fly's medulla (Single and Borst
1998, reviewed by Laughlin 1999). One important function
of these lobula plate tangential cells is to provide
input information for course control (Hausen and
Wehrhahn 1983, Hausen and Egelhaaf 1989, Krapp et al.
1998). A distinct advantage of these cells in this
preparation is that they allow long and stable
recording. When care is taken to do minimal damage to
the fly, and to feed it regularly, the same cell can be
recorded from for several days on end. This is important
in many of our studies of neural coding because there
the general aim of the experiment is to characterize
probability distributions of stimulus response
relations, rather than only averages.

\begin{figure}[tbp]
\begin{center}
\includegraphics[keepaspectratio,width=1.0\textwidth]
{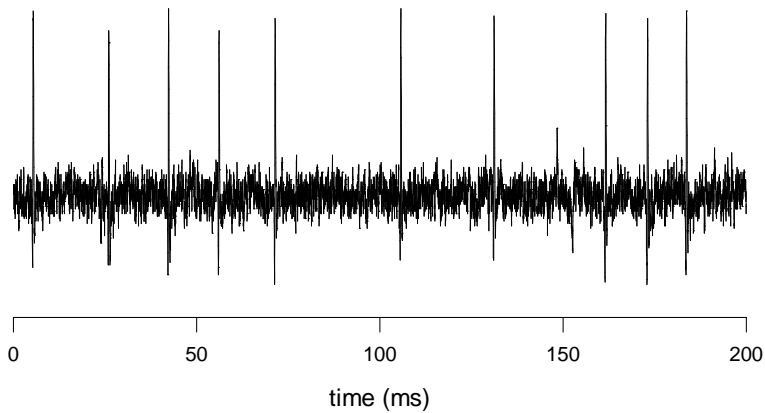}
\caption[ ]{\footnotesize 
  Example of a 200~ms segment of spike activity
  recorded extracellularly from H1. The spikes shown here
  are timed at 10$\mu$s precision and stored for
  off line analysis.}
\end{center}
\end{figure}

The data we present here are all obtained by
extracellular recording with conventional techniques
(see de Ruyter van Steveninck and Bialek 1995 for more
experimental details). The nature of the signal is
drastically different from what we saw before. We are
now dealing with spikes, as depicted in Fig.~13, instead
of analog voltages, and one of the important issues is
how we must interpret their temporal sequences. Spikes
are transmitted by the nervous system at rates of
usually not more than a few hundred per second, and
certainly not at rates typical for photons entering
photoreceptors. In contrast to photons, spikes are
placed on the time axis, one by one, by physiological
processes. It is therefore not unnatural to think that
their position likewise could be read out and
interpreted by other physiological processes. Indeed,
one long standing issue in understanding neural coding
is whether this is the case: Does the timing of
individual spikes matter, or can we afford to coarse
grain time and average spike counts over large windows?
Here we will address that question for the case of H1, a
motion sensitive neuron in the fly's brain. It is
sensitive to horizontal inward motion (see also the
legend to Fig.~3), and its visual field covers just
about a full hemisphere.

In most of the experiments the fly watches an extended
pattern generated on a display oscilloscope (Tektronix
608), written at a 500~Hz frame rate. One drawback of
this setup is that the stimulated area of the visual
field is only a fraction of the total field of the cell,
and that the light levels are rather low, corresponding
roughly to those at dusk. At the end of this chapter we
will present data on a fly that was rotated around the
vertical axis by a computer controlled stepper motor. Doing this 
experiment outside in a wooded environment the fly is 
stimulated with natural scenes of high light intensity, and by 
playing a rotational motion sequence derived from measured flight
data, shown in Fig.~1, one expects to get closer to the ideal of 
presenting naturalistic stimuli.

\subsection{Retinal limitations to motion estimation}
\label{RETLIM}

As we have seen earlier, flies are quick and
acrobatic, so it seems entirely reasonable to assume
that the components of the flight control system are
optimized to perform as accurately and quickly as
possible. This should then obviously be true for the
lobula plate tangential cells as well. Of course some
general principles apply, and, like all sensory neurons
they must work within the limits set by the reliability
of their input signals.

It is instructive to make a rough
estimate of what precision we can expect from a wide
field cell that takes as input photoreceptor signals with
realistic amounts of noise.
\index{motion estimation!retinal limitations} Let us
try to compute an estimate of the limits to timing
precision of the response with respect to stimulus, as
set by the photoreceptor signal quality. This is
relatively easy to do, and it is relevant in a
discussion of coding by spike timing. In many
experiments we stimulate the fly with a large, high
contrast bar pattern that moves randomly, jittering back
and forth in the horizontal direction. The power density
spectrum of the signal we, and presumably H1, are
interested in is the velocity power density, given by
$S_{\mathrm{{vel}}}(f)$. This has dimensions
$(\mathrm{^{\circ}/\mathrm{s}})^2/\mathrm{Hz}$, because
we are dealing with angular velocity, and this is
customarily given in $\mathrm{^{\circ}/s}$.

The photoreceptors in the fly's retina have a profile of angular
sensitivity that is determined by the optics of the lens and
the waveguide behavior of the receptor cell itself; this profile is
often
approximated by a Gaussian. For the blowfly frontal visual field, its
halfwidth is
$\approx 1.4^{\circ}$ (Smakman et al. 1984),
corresponding to a ``standard deviation''
$\sigma_{\mathrm{PSF}}\approx 0.5^{\circ}$. Now suppose
we have a contrast edge with intensity stepping from
$I=I_0\cdot(1+c_0)$ to $I=I_0\cdot(1-c_0)$ aligned on
the optical axis of this Gaussian point spread function.
Then, if the edge moves by a small amount $\delta x$,
the contrast step in the photoreceptor is:

\begin{equation}
\label{eq:DELTAC}
\delta c=\frac{\delta I}{I_0} \approx
\frac{1}{I_0} \cdot \frac{2 \cdot I_0 \cdot c_0\cdot
\delta x} {\sqrt{2\pi}\cdot\sigma}= \frac{c_0 \cdot
\delta x}{\sqrt{\pi/2}\cdot\sigma},
\end{equation}

\noindent which converts a position change $\delta x$
into a contrast change $\delta c$. This allows us to
derive a contrast power spectrum, if we can convert the
velocity power spectrum into the appropriate position
power spectrum. Position is the integral of velocity,
which means that we must divide the Fourier transform of
velocity by frequency to get the Fourier transform of
the position signal (see Bracewell 1978). Here the
relevant quantities are power spectra, so we must use
$f^2$ to make the correct conversion:
$S_{\mathrm{pos}}(f)=S_{\mathrm{vel}}(f)/f^2$. The
contrast power density spectrum is now:

\begin{equation}
S_{c}(f)=\left[\frac{\delta c}{\delta x}\right]^2 \cdot
S_{\mathrm{pos}}(f)=
\frac{2 \cdot c_0^2 \cdot S_{\mathrm{vel}}(f)}
{\pi \cdot \sigma^2 \cdot f^2}.
\end{equation}

\noindent In the experiment we stimulate a large number,
$M$, of photoreceptors and when a wide field pattern
moves rigidly then all photoreceptors are stimulated
in a coherent way. That means that
the total power of the signal available in the
photoreceptor array scales as $M^2$.
Finally, in the experiment we control the velocity
stimulus, and thus $S_{\mathrm{{vel}}}(f)$. All this
combined leads to a contrast signal power spectrum:

\begin{equation}
\label{eq:CTRPSP}
S_{c{_M}}(f)= \frac{2 M^2 \cdot c_0^2 \cdot
S_{\mathrm{vel}}(f)} {\pi \cdot \sigma^2 \cdot f^2},
\end{equation}

\noindent
for the set of $M$ photoreceptors stimulated by
coherent, that is, rigid, motion.

To derive a limit of timing
precision, or equivalently a limiting frequency, we
must compare this available signal spectrum to the
relevant noise power spectrum. In
section~\ref{CTRF_ECN} we defined the equivalent contrast
noise power of a single photoreceptor cell. In a
pool of $M$ photoreceptor cells,  their
independent noise powers add, and we have:

\begin{equation}
N_{c_{M}}(f)=M \cdot N_c(f).
\end{equation}

\noindent
To define a limit to time resolution we
determine the frequency at which 
$S_{c_{M}}(f)/N_{c_{M}}(f)$, the signal to noise
ratio, crosses one:

\begin{eqnarray}
\label{eq:MOTIONSNR}
\frac{S_{c_{M}}(f)}{N_{c_{M}}(f)}
=
\frac{2 M^2 \cdot c_0^2 \cdot S_{\mathrm{vel}}(f)}
{\pi \cdot \sigma^2 \cdot f^2} \cdot
\frac{1}{M \cdot N_c(f)}=\frac{M \cdot S_c(f)}{N_c(f)}.
\end{eqnarray}

\begin{figure}[tbp]
\begin{center}
\includegraphics[keepaspectratio,width=1.0\textwidth]
{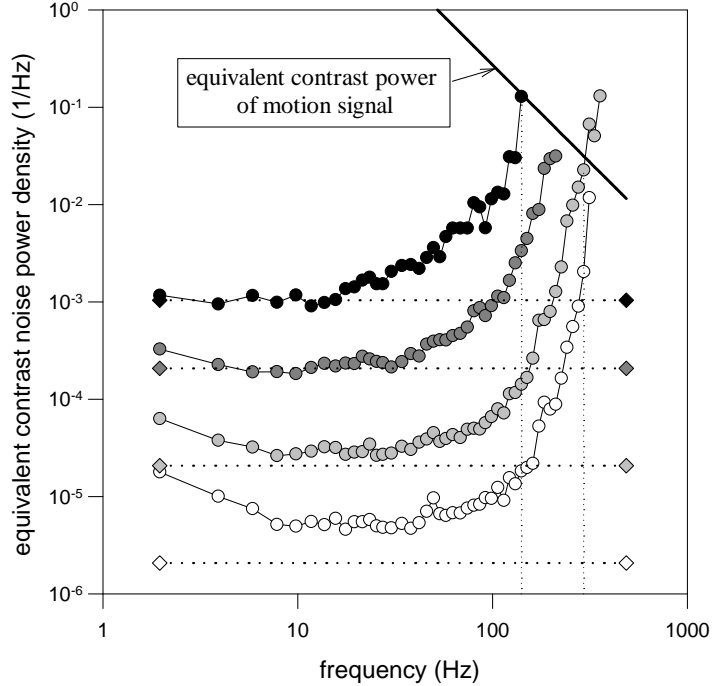}
\caption[ ]{\footnotesize 
  Circles: Equivalent contrast noise power
  spectral density of blowfly photoreceptors at different
  light intensities. The input light intensities,
  expressed in extrapolated bumps per second, are
  indicated by the diamonds connected by dotted lines.
  Solid thick line: Equivalent contrast noise power
  density of a motion signal calculated for a wide field
  pattern, under conditions typical for our
  experiments. See text for further explanation.
}
\end{center}
\end{figure}

\noindent In Fig.~14 we therefore
plot $N_c(f)$ at four light levels for a single
photoreceptor, along with $M \cdot S_c(f)$, for
conditions typical of our experiment: $c_0=0.3$, $M
\approx 3800$, $S_{\mathrm{vel}}(f) \approx 10
(\mathrm{^{\circ}/s})^2/\mathrm{Hz}$, and a small
($\approx 0.3$) correction for the fact that the edge
can not be expected to be exactly at the center (the
correction factor averages over edge position). The
crossover frequency, $f_{\mathrm{cross}}$ of signal to
noise ratio can be read from the figure directly, and
lies between approximately 150 to 400~Hz, depending on
light intensity. That means that when we stimulate H1
with the typical stimuli described here, we may expect
to observe timing precision $\delta t \approx 1/(2\pi
\cdot f_{\mathrm{cross}})$ in the order of one to
several milliseconds if the fly's brain effectively uses
all the motion information present in the photoreceptor
array. We should note that the approximation we make
here is for small $\delta x$ in Eqn.~\ref{eq:DELTAC}.

\begin{figure}[tbp]
\begin{center}
\includegraphics[keepaspectratio,width=1.0\textwidth]
{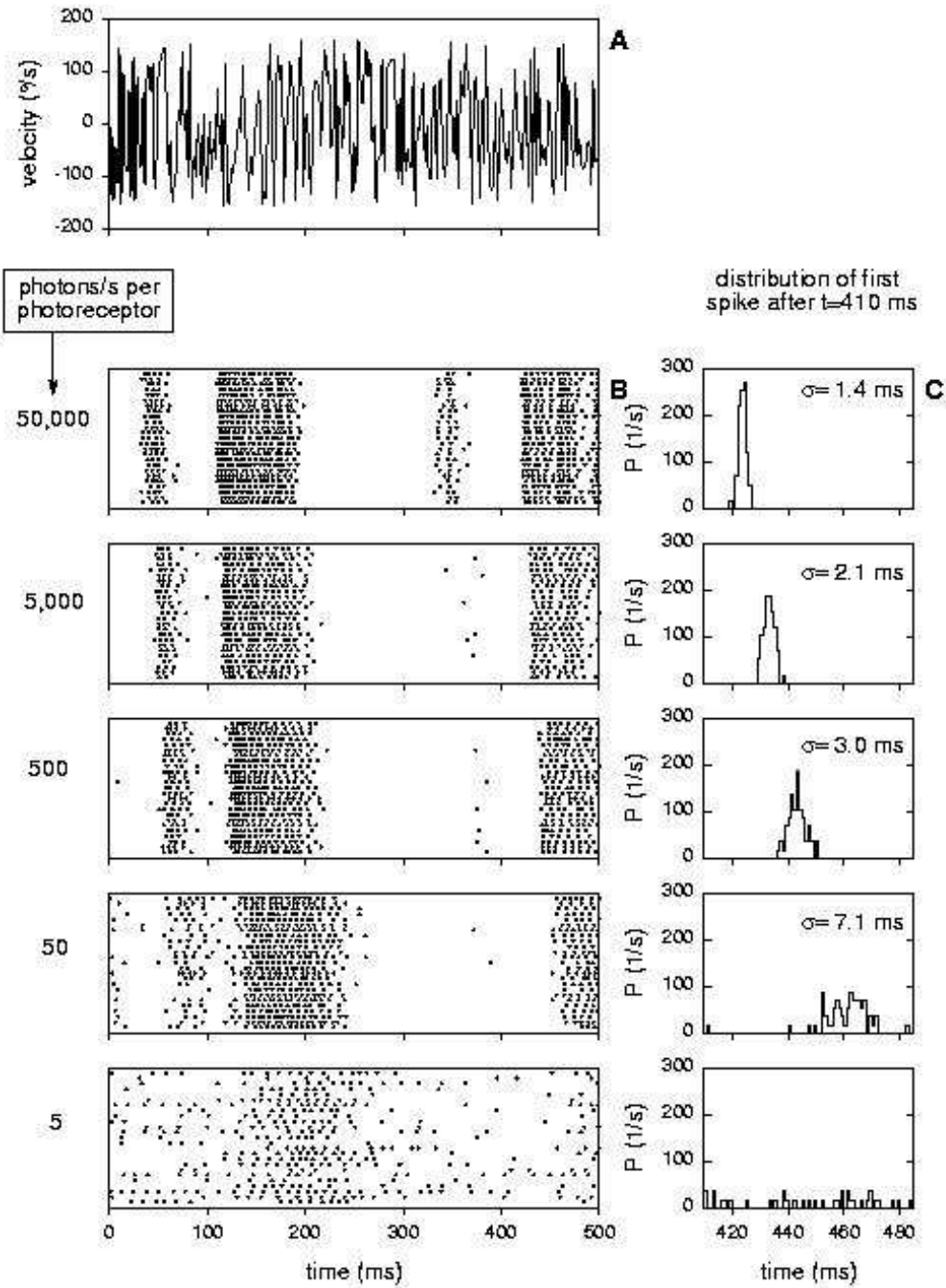}
\caption[ ]{\footnotesize 
  Responses of a blowfly
  H1 neuron to movement of a wide field pattern at
  different mean light levels. In this experiment the fly
  looked at a moving pattern through a round diaphragm
  with a diameter corresponding to 30 horizontal
  interommatidial angles. The velocity was random with
  a flat power spectral density. A: A 500 ms sample
  segment of the stimulus velocity. B: Raster plots
  of H1 spikes obtained at five different light
  intensities given by the estimated average photon flux
  for each single photoreceptor. C: Histograms of
  the timing of the first spike fired after $t=410$~ms.
  With decreasing photon flux the response latency
  increases. Moreover, the peaks become wider (with
  $\sigma$ the standard deviation of a fitted Gaussian),
  suggesting that timing precision may be limited by
  photoreceptor noise. At the lowest light level shown
  here there is still a visible modulation of H1's rate,
  but the timing of a single event is too spread out to
  produce a clear peak.}
\end{center}
\end{figure}

In Fig.~15 we show
raster plots of a 500~ms segment of responses of
H1 to a repeated dynamical motion trace.  The same
experiment was done at different light
intensities, ranging over 4 orders of
magnitude, from an estimated 5 to $5\cdot 10^4$
transduced photons per photoreceptor per second. It is
clear by visual inspection that spikes tend to line up
better when the light intensity is higher, up to the
highest light intensity used in the experiment,
suggesting that external noise may be the limiting
factor (see also Fermi and Reichardt 1963). That
impression is confirmed by the histograms of spike
arrival times to the right of the rasters, which have
standard deviations ranging from 1.4 to 7.1~ms. \index{precision!of
spike timing in H1} These
histograms describe the probability distributions of the
first spike that follows $t=$410~ms. The highest light
intensity in this experiment corresponds roughly to the
second-highest (light gray dots) intensity in the
photoreceptor data of Fig.~14.
At five per second, the lowest bump rate shown here,
bumps in a single photoreceptor typically are
nonoverlapping, and we see that there is still a
modulation of the response of H1. Dubs et al. (1981)
showed that flies respond behaviorally to moving
patterns at light levels where single photon absorptions
are nonoverlapping, and from the classical work of Hecht
et al. (1942) we know that humans can perceive light
flashes under these conditions.

One can think of other
parameters likely to affect the quality of
the input signal for a wide field motion sensitive cell,
for example contrast, field of view, and
stimulus velocity amplitude. In experiments where these
parameters are varied we see effects on spike jitter
qualitatively similar to what is shown here. It thus
seems that the precision of spike timing in H1 is close
to being determined by the information available to it
in the photoreceptor array, i.e., bump latency jitter
sets the threshold under the photon capture rate
conditions of the experiments in Fig.~15. From Fig.~14
we can also read that the photoreceptor equivalent
contrast noise has a very steep frequency dependence at
high frequencies. This means that once the conditions of
the experiment are such that the output accuracy is in
the millisecond regime, only relatively large changes in
stimulus parameters will lead to appreciable
improvements in timing precision.

We can ask a
related question, namely, how reliably can
the arrival time of a spike tell us something about the
strength of the stimulus that preceded it, and how
close does that get to the photoreceptor
limits. If we know what message the neuron encodes, we
can use a computational model that retrieves that
message from the sensory periphery. The model relevant
for H1 is the Reichardt correlator model (Reichardt
1961, Reichardt and Poggio 1976), which describes a
specific functional computation for extracting motion
information from an array of photoreceptors.
\index{Reichardt correlator!and noise} Its basic
interaction is a multiplication of filtered signals
originating from neighboring directions of view. The
Reichardt model was formulated heuristically, but it
has been shown to be the optimal solution to a general
problem of motion estimation in the presence of noise,
in the limit where the signal to noise ratio is low
(Potters and Bialek 1994).

The experiment to measure reliability is very simple (de
Ruyter van Steveninck and Bialek 1995). We present the
fly with a pattern that makes sudden motion steps of
several sizes, and record the responses of H1. From a
large number of presentations we obtain a
histogram of arrival times of the first and the
second spike following the stimulus. If we compare two
such histograms in response to two different step
sizes we can compute the discriminability of those
two stimuli as a function of time. This is done by
framing the question as a decision problem (Green and
Swets 1966). A stimulus is presented once, and the cell
generates a response which is observed by some
hypothetical observer. The question then is: Given the
cell's response, what can the observer say
about the identity of the stimulus, and how often is
that assessment right? This is meaningful only
if the distributions of responses to
different stimuli are known to the observer, and are
also different. In the analysis presented here, the
hypothetical observer can judge the spike train in real
time, starting at the moment of stimulation (or rather
15~ms after that, to minimize effects of spontaneous
rate). The assumption that the time of stimulus
presentation is known does of course not
correspond to any natural situation, as there the timing
of stimuli must be inferred from the sensory input as
well. It is, however, still a valid characterization of
the precision with which the fly's visual brain
performs a computation.

\begin{figure}[tbp]
\begin{center}
\includegraphics[keepaspectratio,width=1.0\textwidth]
{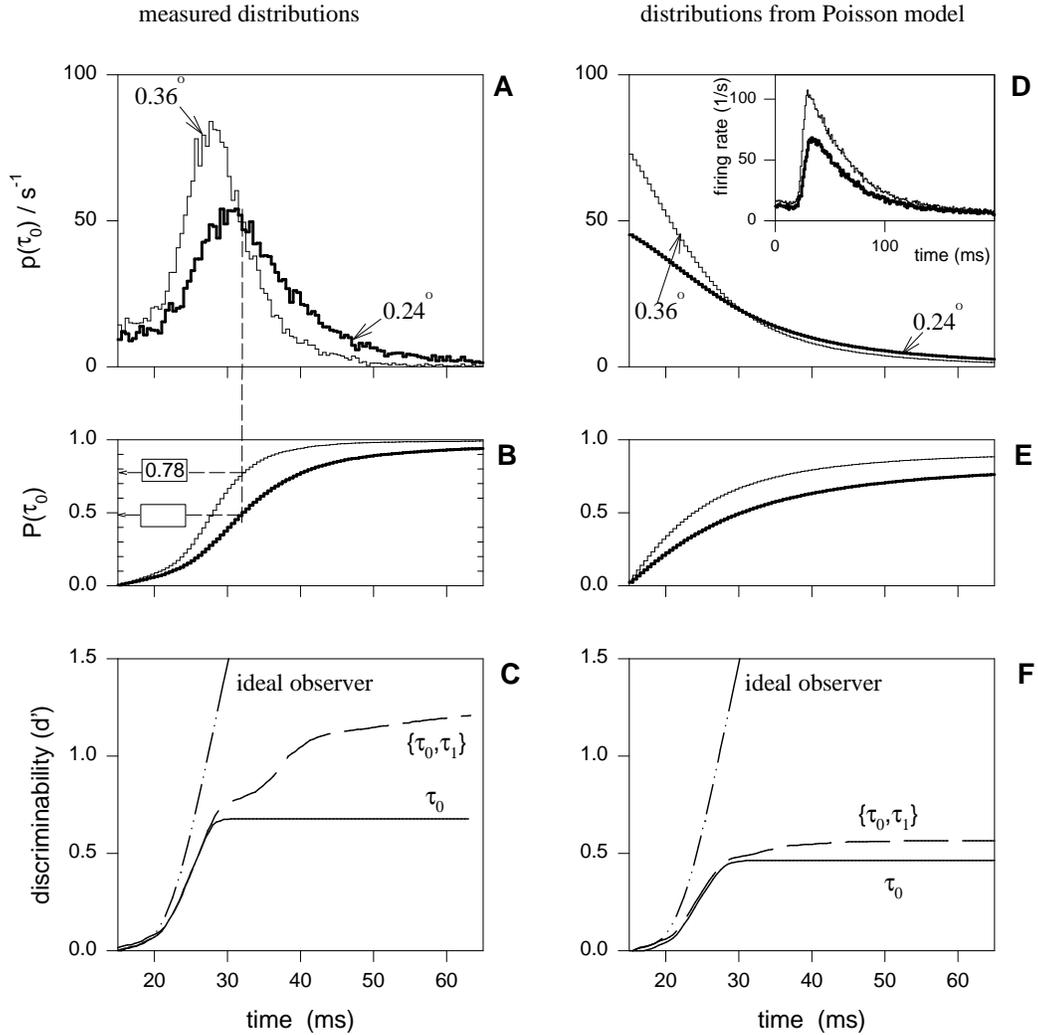}
\caption[ ]{\footnotesize 
 Statistics of H1's responses to
 small motion steps of a wide field pattern, and
 equivalent Poisson statistics. A: Histograms,
 normalized as probability densities of the first spike
 fired starting 15~ms after the stimulus step, for a
 $0.24^{\circ}$ (thick line) and a $0.36^{\circ}$ (thin
 line) step. B: Cumulative distributions for the
 same data as in panel A. C: Time dependent
 discriminability parameter, $d'(t)$, computed for the
 step size pair depicted in A, both for single
 spikes ($\tau_0$), and for spike pairs
 $(\{\tau_0,\tau_1\})$. Also shown is the
 discriminability computed for an ideal observer using
 realistic photoreceptor signals as inputs. D:
 First spike histograms, as in A, but now for a
 modulated Poisson process. These data were computed
 using the measured PSTHs (inset in D) for the two
 step sizes as the rate of the underlying modulated
 Poisson process. E and F: As B and
 C, but now for the modulated Poisson case.}
\end{center}
\end{figure}

Fig.~16a shows two conditional distributions,
$p(\tau_0|S=0.24^{\circ})$ and
$p(\tau_0|S=0.36^{\circ})$ for the first spike arrival
after stimulus presentation, for two step stimuli of
different size. For the large step, the first spike
tends to come earlier than for the small step. This
means that the observer should choose the large step
when he or she sees a short interval, and the small step
for a relatively large interval. The crossover for this
case is at about 32~ms. How accurate will this judgement
be on average? As can be read from the cumulative
probability distributions in Fig.~16B, the probability
for having a spike before 32~ms with the large step is
0.78. Thus if the large step was presented then the
observer will make the right choice with a probability
of 0.78. The cumulative distribution for the small step
at 32~ms is 0.48, which means that if the small step was
presented, the observer will choose correctly in a
proportion of $1-0.48=0.52$. If each step has a prior
chance of 0.5, then in an experiment where the two steps
would be mixed at random, the proportion of correct
decisions would be $P_C=0.5\cdot0.78+0.5\cdot0.52=0.65$.
It is convenient to translate $P_C$ into a distance
measure, and Green and Swets (1966) propose to use the
distance between Gaussian distributions of unit standard
deviation that would give rise to the same value of
$P_C$. This measure is known as $d'$ and is very widely
used in the psychophysical literature.  There is a
simple 1:1 correspondence between $d'$ and $P_C$,
and for our case we find $d'=0.68$. From the given spike
timing distributions we can also construct a continuous
function $d'(t_{\mathrm{obs}})$ that describes the
equivalent distance as it evolves over the observation
time interval since stimulus presentation. To see that,
we should simply divide the distribution in a part that
is described by the measured distribution up to
$t_{\mathrm{obs}}$, assign the probability that is
as yet ``unused'' to one total remaining probability, and
then treat this constructed distribution in the same way
as the previously defined spike timing distributions.
The time dependent $d'(t)$ based on the first spike only
is shown as the solid line in  Fig.~16C. This line
plateaus at $t=32$~ms, as from that moment on
the choice will always be fixed. Instead of considering
only the first spike we can also look at the combination
of the first and the second spike arrival time, that is
$\{\tau_0,\tau_1\}$. The distributions for these are not
shown, but the reasoning is entirely similar. The end
result, $d'$ as a function of time after presentation,
is shown in the same panel as the dashed line. It is
clear in this case that the second spike carries
substantial extra information about stimulus identity.

The comparison of the measured data to the ideal motion
detector model can now be made. We measure
representative photoreceptor power spectral
densities, and the number of photoreceptors stimulated
in the H1 experiment, and then apply the Reichardt model
to compute its average step response as well as its
output power spectral density (de Ruyter van Steveninck
and Bialek 1995). From these we compute a time
dependent $d'_{\mathrm{model}}(t)$, which is plotted as
the dash-dot line in  Fig.~16C. The
crucial comparison is between the slope of the measured
and the computed $d'(t)$. In the range of 23-28~ms,
$d'_{\mathrm{model}}(t)$ rises about twice as fast as
the measured $d'(t)$. This shows that H1 approaches,
within a factor of two,
the performance of an optimal motion
detector limited only by noise in the
photoreceptor array. \index{motion
detection!efficiency of} \index{efficiency!of
motion detection} Given that the signal passes through
at least four synapses to be computed, this precision is
quite remarkable. In this case we measure how accurately
H1 represents the amplitude of motion steps. The
estimation takes place over a somewhat extended time
interval, and is therefore not limited by the bump
jitter that sets the timing resolution of the
photoreceptor array, but rather by low frequency
(roughly below 100 Hz) accuracy of the photoreceptor.  As
this latter is close to the photon shot noise limit
(Figs.~7,14) we are reminded that the precision of
neurons in a functioning brain is not just given by the
physiology, but is determined in part, or in this case
maybe even dominated, by the statistical properties of
the stimulus (see also Bialek et al. 1991, de Ruyter van
Steveninck and Bialek 1995, Rieke et al. 1997).

Motion discrimination using the spike train output of H1
thus provides us with an example in which the
performance of the nervous system approaches basic
physical limits set by the structure of the inputs in
the retina. There are other examples of this
near optimal performance, in systems ranging from human
vision to bat echolocation to spider thermoreception
(for discussion see Rieke et al. 1997). This level of
performance requires the nervous system to meet two very
different requirements. First, the system must be
sufficiently reliable or ``quiet'' that it does not add
significant excess noise. This is especially challenging
as the signals propagate through more and more layers of
neurons, since the synapses between cells are sometimes
observed to be the noisiest components. Second, optimal
performance requires that the system make use of very
particular algorithms that provide maximal separation of
the interesting feature (motion, in the case of H1) from
the background of noise. In general these computations
must be nonlinear and adaptive, and the theory of
optimal signal processing (Potters and Bialek 1994)
makes predictions about the nature of these
nonlinearities and adaptation that can be tested in
independent experiments. \index{optimal signal
processing} \index{Poisson statistics!and neural coding} 

In our broader discussion on the relevance of Poisson
firing it is now interesting to quantify to what extent
deviations from Poisson behavior help in encoding the
stimulus in a spike train. To get an idea of this we do
exactly the same analysis as described above, but then
on synthetic spike trains that are generated by a
modulated Poisson process with the same time dependent
rate as the measured spike train. The inset in Fig.~16D
shows the post stimulus time histogram for the two step
sizes used before in the analysis. From these two, the
spike arrival distributions in panel d are computed, and
from these again we construct $d'(t)$. Figure 16F shows
the end result: The discriminability based on timing of
the first spike alone goes from $d'=0.68$ to $d'=0.46$,
while in the Poisson case the second spike adds only 0.1
to the $d'$ based on the first, so that at 50~ms after
the step, the Poisson value for $d'$ is less than half
that measured from real spikes. Neural refractoriness
was not incorporated in the synthetic train, and it
seems likely that that the increased reliability of the
real neuron can at least be partly attributed to that. 

\subsection{Taking the fly outside: Counting and
timing precision in response to natural stimuli}
\label{FLYOUTSIDE}

There is a long tradition of using discrimination 
tasks, as in the step discrimination experiment of the 
previous section, to probe the reliability of perception
in humans and also the reliability of neurons.  But such
simple tasks are far from the natural ones for 
which evolution selected these neurons.  As a fly flies 
through the world, angular velocity varies continuously,
and this variation has a complicated dynamics.  In the 
past decade, a number of experiments has been done 
which attempt to approach these more natural 
conditions.\index{motion stimuli!natural}\index{natural 
stimuli!motion} Specifically, experiments with 
pseudorandom velocity waveforms presented 
on display oscilloscopes have been used to measure information
transmission in H1 by reconstructing the stimulus from the spike train
(de Ruyter van Steveninck and Bialek 1988, Bialek et al. 1991, Haag and
Borst 1997), and by more direct methods (de Ruyter van Steveninck et al.
1997, Strong et al. 1998). The general conclusion of these studies is
that the timing of spikes in the millisecond range does indeed carry
significant information, in line with the photoreceptor limits discussed
earlier. One would like to know to what extent such conclusions are also
relevant to still more natural conditions encountered in fly 
flight (see for example Warzecha et al. 1998).

Ideally one would perform experiments in the natural
habitat of the animal, while it is behaving as freely as
possible. Good examples are the study of responses of
auditory neurons in \emph{Noctuid} moths to the cries of
bats flying overhead (Roeder 1998), and the study of
optic nerve responses of \emph{Limulus} lateral eye
while the animal is moving under water (Passaglia et al.
1997). Of course, in each specific case concessions are
made to be able to record neural signals, and one must
decide what is the best compromise between realistic
conditions and getting interpretable data.

Here we present data from a setup that allows us to
record from a fly while it is rotating on a stepper
motor. The rotational motion is mechanically precise,
and arbitrary rotation sequences can be programmed. In
the case described here the fly was rotated with a time
sequence corresponding to the rotations executed by the
flies shown in Fig.~1 (based on Land and Collet 1974),
except that for stability reasons the amplitude of the
entire trace was set to half the value measured from the
real flies. The flight trajectories were repeated with
their sign inverted, so that for each trajectory we
stimulate H1 in complementary ways. It is thus as if we
record from the two H1 cells at opposite sides of the
head. The setup is portable and the experiment was done
outside in the shade of some bushes on a sunny
afternoon. Therefore, the light intensities, the
stimulated area of the visual field, and the spatial
characteristics of the scene are realistic samples of
what the animal encounters in nature. The motion trace
is somewhat natural, although the rotational velocities
are smaller than those measured in free flight, and the
measurement was done on a different species of fly.

Important for our analysis is that we can repeat the
same motion trace a large number of times, which is of
course not really a part of natural behavior. However,
it allows us to make quantitative statements about
information transmission in the measured spike trains
that rely only on the degree of
reproducibility of the response to repeated
stimuli. Because of this, those statements are
independent of any assumptions on how the stimulus is
encoded in the spike train,
so in that sense they are rather universal.

\begin{figure}[tbp]
\begin{center}
\includegraphics[keepaspectratio,width=0.9\textwidth]
{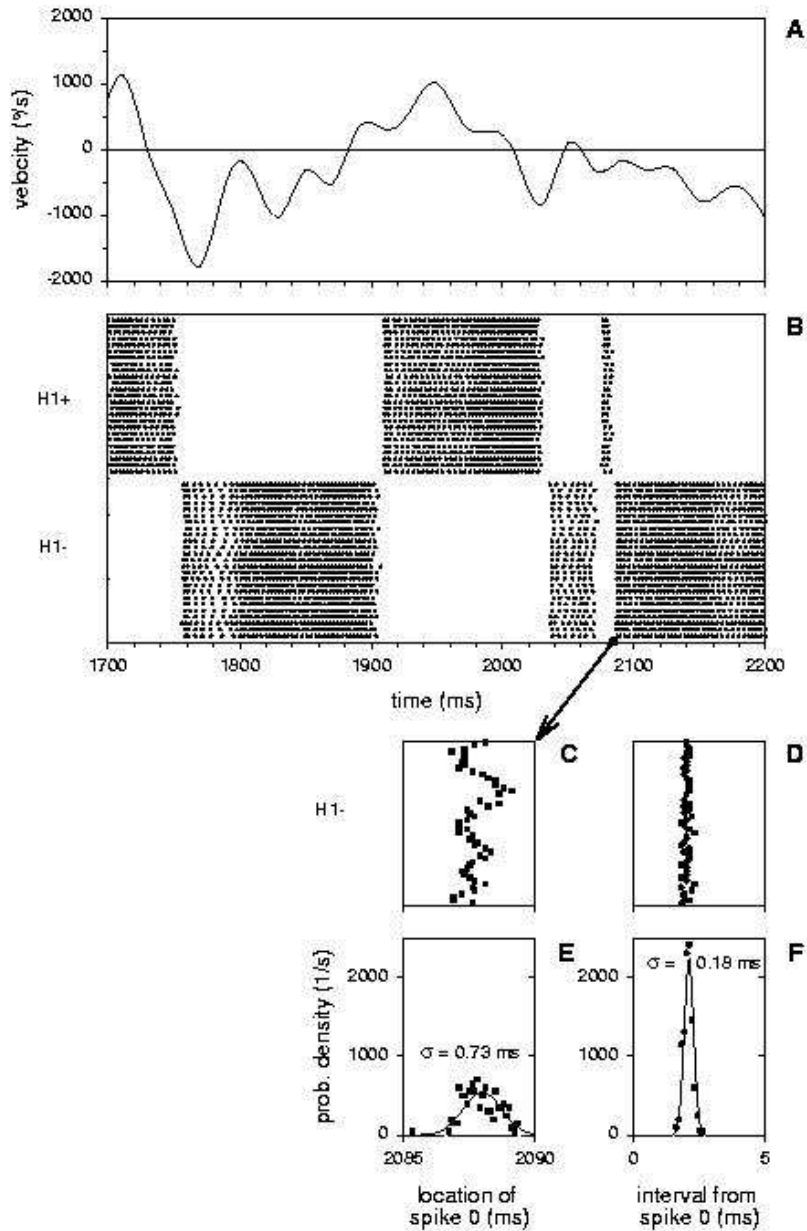}
\caption[ ]{\footnotesize 
  Direct observations of H1 spike
  timing statistics in response to rotational motion
  derived from Land and Collett (1974) free flight data
  (see Fig.~1). The fly was immobilized in a specially
  designed miniature recording setup, which was fixed to a
  computer controlled stepper motor. This setup was used
  in a wooded outdoor environment, in which the fly was
  rotated repeatedly (200 times in total) along the same
  motion trajectory. For technical reasons the rotational
  velocity used in this experiment was scaled down to half
  of the free flight value. A: A 500 ms segment of
  the motion trace used in the experiment. B: Top:
  raster of 25 trials showing occurrences measured from
  H1. Bottom: 25 trials with spike occurrences from the
  same cell, but in response to a velocity trace that was
  the negative of the one shown in A. For ease of
  reference we call these conditions H1+ and H1-
  respectively. C: 25 samples of the occurrence time
  of the first spike fired by H1- following $t$=2085~ms
  (indicated by the arrow connecting the axis of panel
  B to panel C). D: Time interval from
  the spike shown in C to the spike immediately
  following it. E: Probability density for the
  timing of the spike shown in C. The spread is
  characterized by $\sigma=0.73$~ms, which is defined here
  as half the width of the peak containing the central
  68.3\% of the total probability. If the distribution
  were Gaussian, then this would be equivalent to the
  standard deviation. Here we prefer this definition
  instead of one based on computing second moments. The
  motivation is that there can be an occasional extra
  spike, or a skipped spike, giving a large outlier which
  has a disproportionate effect on the width if it is
  calculated from the second moment. Filled squares
  represent the experimental histogram, based on 200
  observations; the solid line is a Gaussian fit. F:
  Probability densities for the same interspike interval
  as shown in D. The definition of $\sigma$ is the
  same as the one in E.}
\end{center}
\end{figure}

Fig.~17 presents data from such an experiment. The
rotational velocity waveform is shown in panel A. Note
that the velocity amplitudes are very large compared to
those of the white noise stimuli shown in Fig.~15. The
traces in panel B are labeled H1$+$ and H1$-$. In reality
they were obtained from the same H1 cell, but with a
switch of sign in the velocity trace, as described
earlier. It is clear that H1$+$ and H1$-$ alternate their
activity quite precisely, and that repeateable patterns
of firing occur, such as the pair of spikes in H1+ at
about 1900 ms. The edges in spike activity are sharp.
Panel C shows the position of the first spike occurring
after 2085~ms in H1$-$. A histogram of arrival times of
this spike is shown in Fig. 17E, together with a
Gaussian fit with standard deviation 0.73 ms.
\index{spike timing precision! and natural stimuli}

In addition to this precision of spike timing with
reference to the stimulus, there can also be an internal
reference, so that the relative timing of two or more
spikes, either from one neuron or among different
neurons, may carry information (MacKay and McCulloch,
1952). \index{spike timing!relative}\index{timing
precision!of spike intervals} Panel D and F give an
example, where H1 generates a 2 ms interspike interval 
upon a particularly strong stimulus with a standard 
deviation of 0.18 ms.

Consequently,
interspike intervals may act as special symbols in the
code, carrying much more information about the stimulus
than what is conveyed by two single spikes in isolation.
This was shown indeed to hold for H1 (de Ruyter van
Steveninck and Bialek 1988, Brenner et al. 2000).
Findings like these should make us cautious. For a
complete description of the spike train, timing
precision at different levels of resolution may be
required, depending on what aspect of the spike train we
are talking about. In particular, relative spike timing
may have to be much better resolved than absolute timing
to recover neural information (see specifically Brenner
et al. 2000).

\begin{figure}[tbp]
\begin{center}
\includegraphics[keepaspectratio,width=1.0\textwidth]
{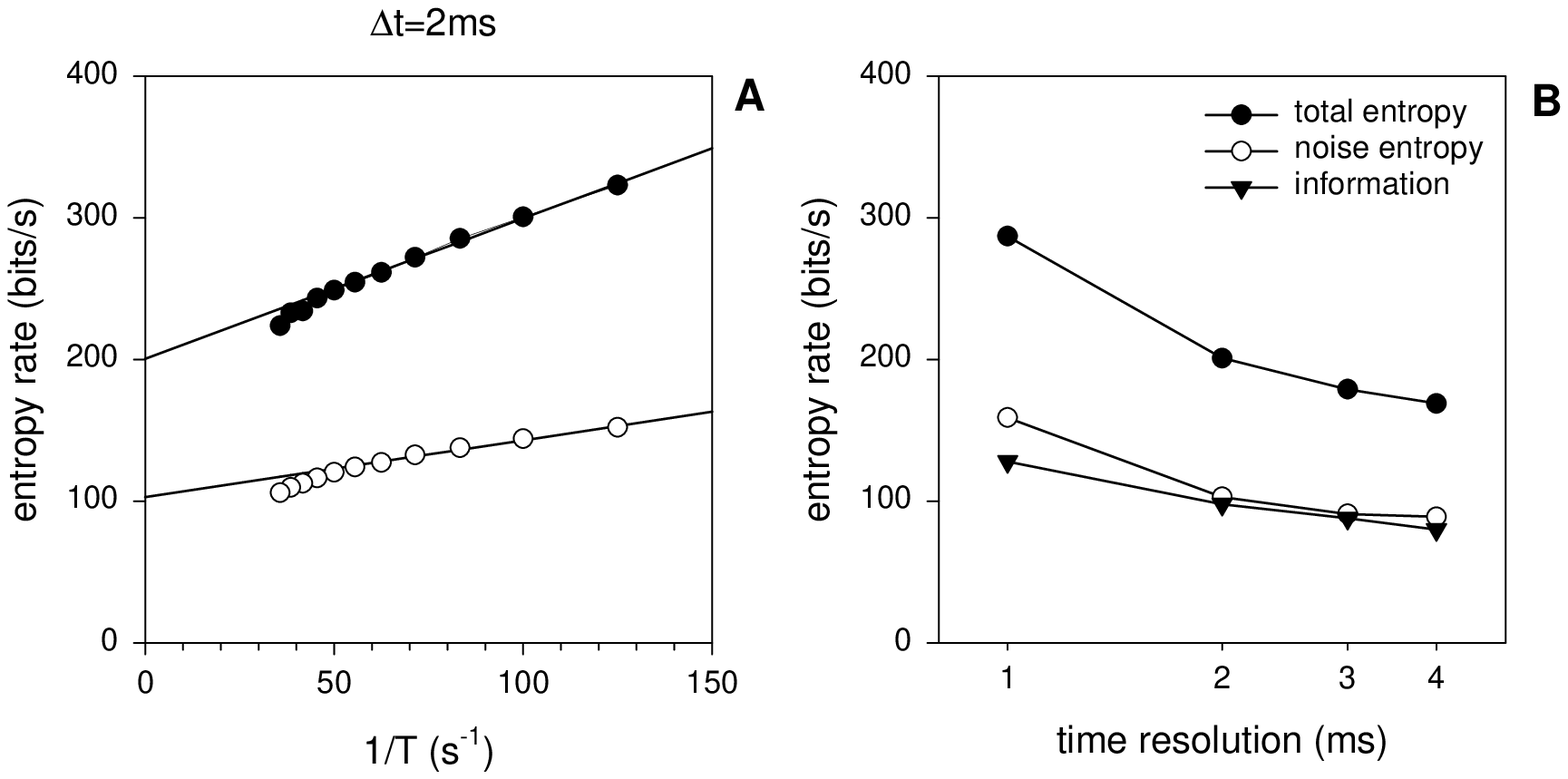}
\caption[ ]{\footnotesize 
  Information in firing patterns obtained from an
  experiment with naturalistic motion stimuli (see legends
  for Figs.~1,~17). A: Rate of total entropy and
  noise entropy as a function of $1/T$, for time
  resolution $\Delta t=2$~ms. The figure shows the
  corresponding entropy rates, that is the values of total
  and noise entropy as defined above, divided by $T$. The
  fits (solid lines) to the two data sets are extrapolated
  to zero value of the abscissa, corresponding to $T \to
  \infty$. The difference of the extrapolated rate for the
  total entropy and the noise entropy respectively, is the
  estimate of the information rate at the given time
  resolution. See Strong et al. (1998) for a more detailed
  explanation. B: Rates of total entropy, noise
  entropy and information transmission were computed as
  explained in the text, and plotted here for different
  values of time resolution. It is clear that even in the
  millisecond range the information transmission increases
  when time resolution becomes finer.}
\end{center}
\end{figure}

The data presented above show
episodes in the stimulus that induce accurately timed
events in the spike train, on the millisecond scale. One
may worry that such events are very special,
and that most spikes are not
well defined on the time axis. In other words, we
need a ``bulk'' measure of spike timing
precision. The most general way of specifying that is
to study the information transmitted by the spike
train as a function of time resolution $\Delta t$. We
do that here by estimating two measures of entropy, the
total entropy and the noise entropy, directly from the
spike train (cf. de Ruyter van Steveninck et al. 1997,
Strong et al. 1998).\index{total entropy}\index{noise
entropy}\index{entropy!of spike trains} Loosely
speaking, the total entropy  measures the size of the
neuron's ``vocabulary.''  We calculate it from the
distribution, $P(W)$, of neural firing patterns, or
words $W$: $S_{\rm total}=-\Sigma_W P(W) \log_2[P(W)]$.
Here $W$ is a vector of $n_W$ entries, and each entry
gives a spike count in a bin of size $\Delta t$. All
$n_W$ bins taken together form a string of length $T$.
Typical values for $\Delta t$ are one to a few ms, while
$T$ is of order 5-30 ms. $P(W)$ is approximated by the
histogram of all firing patterns that the neuron
generates in the experiment. \index{firing
patterns!statistics} The noise entropy characterizes how
much the neuron deviates from repeating the same firing
pattern at the same phase of a repeated stimulus. If the
stimulus is periodic in time with period $T_{\rm stim}$,
then from an experiment with a large number of
repetitions we can form histograms of firing patterns
$W(t)$ at each time $t$, $0\leq t \leq T_{\rm stim}$. If the
neuron were an ideal noiseless encoder then the response
would be the same every time, and for each presentation
we would find the same firing pattern $W(t)$ at time 
$t$. Then $P(W|t)$ would equal one for $W=W(t)$,
and zero otherwise, so that the  noise
entropy would be zero. In practice, of course, the noise
entropy differs from zero, and it will also vary with
time $t$. The time average of the noise entropy is
$S_{\rm noise}=\overline{S_{\rm noise}(t)}=\overline{-\Sigma_W
P(W|t) \log_2[P(W|t)]}$. The information
transmitted by the neuron is the difference of
the total and noise entropies, and therefore depends on
both $T$ and $\Delta t$: $I(T,\Delta
t)=S_{\rm total}-\overline{S_{\rm noise}}$ (Strong et al. 1998).
With enough data we can extrapolate to the limit $T
\to \infty$ to get an estimate of the information
rate. \index{information transmission!in spike train} If
we quantify that limiting information rate as a function
of the time resolution $\Delta t$, we finally arrive at
a reasonable bulk measure of the time resolution at
which information can be read out from the neuron.

The results are shown in Fig.~18B, which plots the total
entropy rate, the noise entropy rate, and the
information rate, all as a function of time resolution.
The information rate still increases going from
$\Delta t=2~$ms to $\Delta t=1~$ms,
to reach 120 bits per
second. The efficiency of encoding, that is the
proportion of the total entropy used for transmitting
information, is about 50\% for $\Delta t\ge2~$ms,
and slightly lower than that for $\Delta t=1~$ms.
Because the information transmission rate must be
finite, and the total entropy grows without bound as
$\Delta t$ becomes smaller, the efficiency must go to
zero asymptotically as $\Delta t$ goes to zero. Due to
the limited size of the dataset it is not possible to
make hard statements about the information
transmitted, and thus the efficiency, at time
resolutions better than 1~ms. It is clear, however, that
spike timing information in the millisecond range, also
under natural stimulus conditions, is present in the
spike train, and that it could be highly relevant to the
fly for getting around.

\begin{figure}[tbp]
\begin{center}
\includegraphics[keepaspectratio,width=1.0\textwidth]
{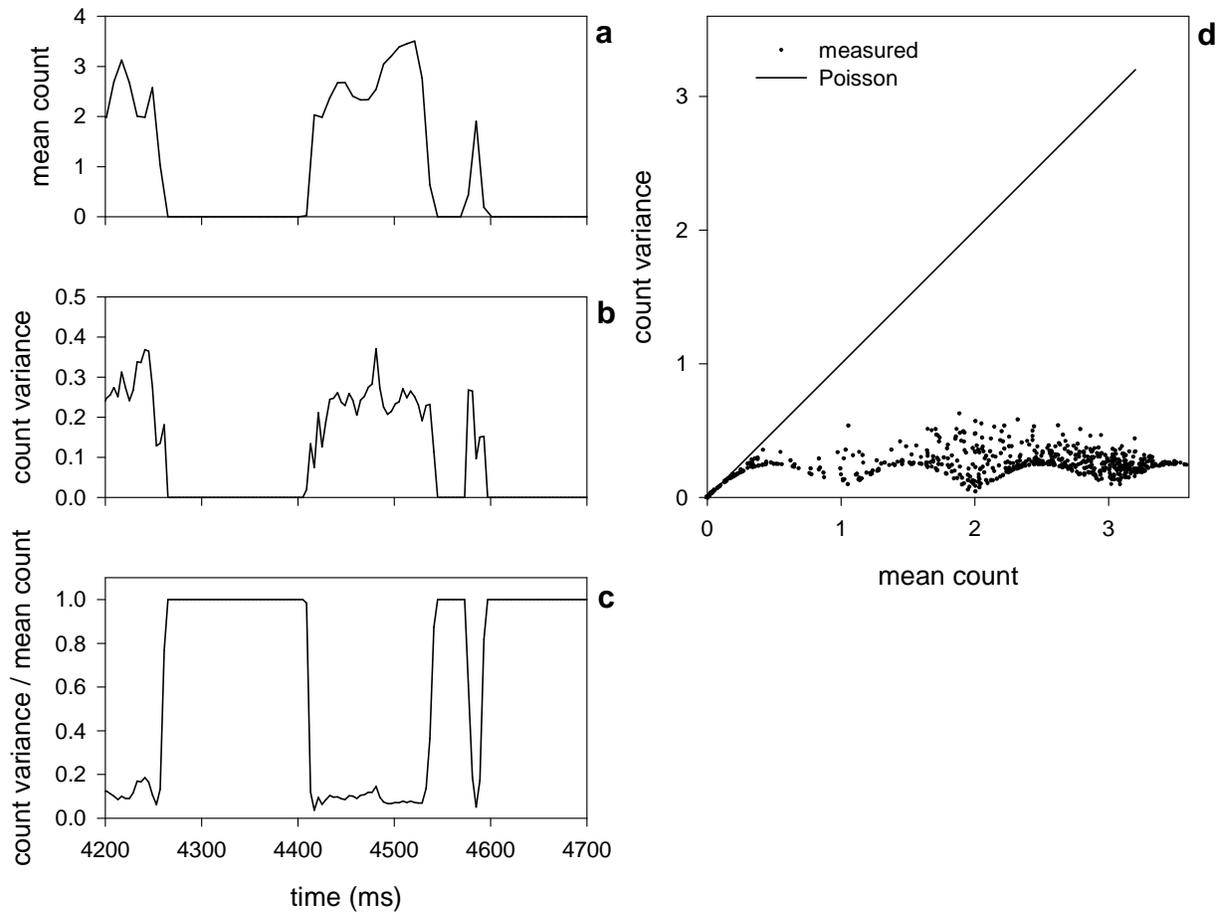}
\caption[ ]{\footnotesize 
  Example of counting statistics in response to
  a natural motion stimulus. The data are for the segment
  represented as H1+ in Fig. 17B. A: Ensemble
  average count in a 10~ms wide sliding windows.
  B: Ensemble variance of the count in the
  same sliding window as in A. C: Ratio of
  ensemble variance to ensemble average. Where both the
  variance and the mean are zero the value of the ratio is
  set to 1. For a Poisson process, all datapoints in this
  plot should have a value of 1. D: Dots: Scatter
  plot of simultaneous pairs of the ensemble variance and
  the ensemble average count. Straight line: Statistics of
  a Poisson process.}
\end{center}
\end{figure}

In the spirit of the discussion in section~\ref{RETLIM} we
can also ask whether under these more natural
conditions, spikes are generated according to a
modulated Poisson process. As mentioned in
section~\ref{POISSDESC}, the variance of a Poisson
distribution is equal to its mean. Furthermore, if
spikes are generated according to a modulated Poisson
process, the spike count in a certain window should
spread according to a Poisson distribution.
\index{Poisson distribution} Thus, if we compare
segments of the responses to repeated identical stimuli,
and we compare the variance of the count in a large
number of such segments, we can see whether the spike
statistics deviate from Poisson (for more details see de
Ruyter van Steveninck et al. 1997). In Fig. 19 we show
the response segment labeled H1+ in Fig. 17B, and we
compute both the mean count and the variance across
trials, for a 10~ms wide sliding window. When we plot
the ratio of variance to mean we see that, as soon as
there is spike activity, the ratio drops to values
between 0.1 and 0.2. The comparison of the scatter plot
to the Poisson behavior in Fig.~19D makes it clear that
there is no strong overall trend for the variance to
scale with the mean for these conditions (this, however,
may be different for much longer time windows, see Teich
and Khanna 1985). Similar results were reported for
other systems (Berry et al. 1997, Meister and Berry
1999) and for fly H1 in laboratory conditions (de Ruyter
van Steveninck et al. 1997).

These results suggest that when stimuli
are dynamic enough, spiking sensory neurons may operate
in a regime far removed from the Poissonlike behavior
they are often assumed to have. This sub-Poisson
behavior can be attributed, at least partly, to the
relative refractoriness of spiking cells. Refractoriness
tends to regularize spike trains (Hagiwara 1954), and
to increase the information carried by short intervals
(de Ruyter van Steveninck and Bialek 1988, Brenner et
al. 2000). Therefore, when operating at the same
time dependent rate, the real H1 neuron carries much
more information about the stimulus than a modulated
Poisson train (de Ruyter van Steveninck et al. 1997).

\section{Discussion and Conclusions}

As the fly moves through its environment, its array of
photoreceptors contains an implicit representation of
motion, and this representation is corrupted by noise.
>From this the visual brain must extract a time
dependent estimate of motion in real time. In some
limiting cases, such as stepwise motion and
the high frequency limit of white noise motion (cf.
section~\ref{RETLIM}), we can estimate the limits to the
precision with which such a running estimate can be made.
In those cases we find that the
computation performed by the fly's brain up to H1 is
efficient in the sense that that H1 retreives a
substantial fraction of the motion information
implicitly present in the photoreceptor array. To
support those conclusions we have presented data
on the statistical efficiency of signals at
several levels within the blowfly's visual system.

Photoreceptors are stimulated by a Poisson stream of
photons and, as long as the modulation is not too
strong, they respond linearly to contrast. At low
frequencies (up to 50-100 Hz, depending on illumination)
the signal to noise ratio with which they encode
contrast is close to the limits imposed by the Poisson
nature of photon capture. Thus, blowfly photoreceptors
are efficient at frequencies up to 50-100 Hz in the
sense that they use almost all the information present
in the photon stream. At higher frequencies the
photoreceptor loses efficiency as latency jitter
becomes the dominating noise source.

The LMCs, directly postsynaptic to the
photoreceptors, are also efficient (Laughlin et al.
1987; this chapter, section~\ref{PRLMCSYN}). This means
that neural superposition indeed works as was suggested
long ago by Braitenberg (1967) and Kirschfeld
(1967). As the LMCs receive their signals through an
array of chemical synapses, the measured signal to noise
ratio of the LMC sets a lower bound to the precision
with which these synapses operate. If we hypothesize
that vesicle release can be thought of as a modulated
Poisson process, then from the measurements we estimate
that each synaptic active zone should emit vesicles at a
tonic rate of well over a hundred vesicles per second.
Based on similar considerations Laughlin et al. (1987)
arrive at an even higher number.
Direct measurements of vesicle release in this system
are unavailable, but recent measurements in goldfish
bipolar cells (Lagnado et al. 1999) give a value six
times lower than our estimates of tonic rates. This
suggests that the hypothesis of Poisson release could be
wrong, and that vesicle release is much more tightly
regulated. Perhaps the large number of different
proteins involved in vesicle docking and release (Kuno
1995), and the delicate anatomical ultrastructure of the
synapse (Nicol and Meinertzhagen 1982, Meinertzhagen and
Fr\"{o}lich 1983) have a role to play in this type of
regulation.

We find that H1 can
generate spikes with strongly subPoisson statistics, if
it is driven by stimuli representative for at least some
of the natural behavior of the fly. Chasing behavior in
flies plays an important role in reproduction and
territorial defense (Land and Collett 1974) and it
presumably taxes the fly's sensory and motor systems to
the fullest. It is therefore an interesting limit in
which to study the performance of the fly's visual
information processing capabilities. However, flies do
not often engage in this behavior, and the chases
typically last not much longer than a second. One may
therefore ask how long the visual system can keep up
with these strong dynamic stimuli. As a casual
observation, it seems that H1 keeps on reporting about
the fly's motion, and we see no signs of habituation
when the fly spins around in simulated flight patterns
for up to 20 minutes.

A fly, its name notwithstanding, spends most of its time
sitting still (Dethier 1976, on page 13 gives a
revealing list of how flies spend their time). If the
fly's environment does not move, or if movement is
steady and slow enough, H1 generates spikes
approximately as a Poisson process. But at the same mean
firing rate, when stimulated with strong dynamic
signals, H1 can be firing far from Poisson, and this
makes the encoding more efficient in the sense that the
information per spike increases substantially (de Ruyter
van Steveninck et al. 1997). Over the years the Poisson
process has been the model of choice for describing
neural firing statistics, at least to first
approximation (see for example Tolhurst et al. 1983,
Britten et al. 1993). As H1 can be either close to that
limit or far from it, depending on conditions, it is a
matter of debate which condition is more relevant.
Reference to natural behavior is not conclusive
here, because both sitting still and chasing other flies
are natural behaviors. We feel that studying neural
responses to dynamic stimuli is more interesting and
rewarding, both because there is already a long history
of characterizing responses to static stimuli, and
because one can reasonably assume that well designed
dynamic stimuli test the information processing
capabilities of the nervous system to the fullest.

From the experiment it has also become clear that H1
can generate spikes that are locked to certain
stimulus events with millisecond timing precision.
Moreover, interspike intervals can be defined even
better, and we saw an example in which an interval was
generated with $\approx$ 0.18~ms accuracy. Overall,
the spike train carries information at
$\approx 50\%$ efficiency at least down to the
millisecond time scale.

At the very least, the combination of these observations
should make one cautious in interpreting or modeling
neural signals as modulated Poisson processes. As we
saw, under some conditions this may be a fair
approximation but in others, specifically those that
approach conditions of natural flight, it definitely is
not. One may wonder whether this latter observation
is more generally valid. Here the report by
Gur et al. (1997) offers an important clue: The authors
studied cells in monkey primary visual cortex, and
noticed that these showed marked sub-Poisson statistics
when care was taken to exclude episodes with eye
movements from the analysis. Thus, from the point of
view of specifying neural reproducibility, eye movements
add variability. But that of course is by no means
necessarily true for the monkey if it knows when it
holds its eyes still. It is thus quite possible that
sub-Poisson firing is a more general phenomenon, and
relevant to natural stimulus conditions; this could
be revealed in experiments designed to control carefully
for variability in the neural response that is knowable
by the animal.

We have encountered an example of this in the H1 data
presented in section~\ref{FLYOUTSIDE}. Fig. 17C shows
the arrival times of spikes with reference to the
stimulus, and these jitter by 0.73~ms. Fig.
17D shows the jitter in the interspike interval
beginning with the spike depicted in panel C, which at
0.18~ms is much tighter. This implies that the 
fluctuations in the absolute timing of the two spikes 
are strongly correlated. If the interval length plays a 
role in the interpretation of H1 somewhere downstream, 
then in a sense the fly corrects for the fluctuations by
knowing that they are correlated. It would even be more
interesting if fluctuations correlated among different
neurons, as that would mean that the relative timing
between spikes from different cells may carry extra
information. Preliminary data from double recording
experiments indicate that this is indeed the case for
the two H1 neurons in the left and right lobula plates,
again under approximately natural stimulus conditions.

\section*{Acknowledgments}

\noindent The work presented here grew out of
collaborations and discussions with many
people, and it is our great pleasure to thank Naama
Brenner, Simon Laughlin, Geoff Lewen, Roland
K\"{o}berle, Al Schweitzer, and Steve Strong for sharing
their insights and helping us out. We thank Simon
Laughlin in particular for his thoughtful comments.

\end{document}